\def\appropto{\mathrel{\vcenter{
  \offinterlineskip\halign{\hfil$##$\cr
    \propto\cr\noalign{\kern2pt}\sim\cr\noalign{\kern-2pt}}}}}

\documentclass[twocolumn]{aastex62}

\received{12 Nov 2018}
\revised{--}
\accepted{--}

\shorttitle{Oxygen column densities in a precipitation-limited CGM}
\shortauthors{Voit}

\begin{document}

\title{\bf Ambient Column Densities of Highly Ionized Oxygen in Precipitation-Limited Circumgalactic Media}

\correspondingauthor{G.\ Mark Voit}
\email{voit@pa.msu.edu}

\author{G.\ Mark Voit}  
\affiliation{Department of Physics and Astronomy,
                 Michigan State University,
                 East Lansing, MI 48824} 

\begin{abstract}
Many of the baryons associated with a galaxy reside in its circumgalactic medium (CGM), in a diffuse volume-filling phase at roughly the virial temperature.  Much of the oxygen produced over cosmic time by the galaxy's stars also ends up there.  The resulting absorption lines in the spectra of UV and X-ray background sources are powerful diagnostics of the feedback processes that prevent more of those baryons from forming stars.  This paper presents predictions for CGM absorption lines (O~VI, O~VII, O~VIII, Ne~VIII, N~V) that are based on precipitation-regulated feedback models, which posit that the radiative cooling time of the ambient medium cannot drop much below 10 times the freefall time without triggering a strong feedback event.  The resulting predictions align with many different observational constraints on the Milky Way's ambient CGM and explain why $N_{\rm OVI} \approx 10^{14} \, {\rm cm^{-2}}$ over large ranges in halo mass and projected radius.  Within the precipitation framework, the strongest O~VI absorption lines result from vertical mixing of the CGM that raises low-entropy ambient gas to greater altitudes, because adiabatic cooling of the uplifted gas then lowers its temperature and raises the fractional abundance of O$^{5+}$.  Condensation stimulated by uplift may also produce associated low-ionization components.  The observed velocity structure of the O~VI absorption suggests that galactic outflows do not expel circumgalactic gas at the halo's escape velocity but rather drive circulation that dissipates much of the galaxy's supernova energy within the ambient medium, causing some of it to expand beyond the virial radius.
\end{abstract}


\keywords{ISM: general --- galaxies: halos --- intergalactic medium --- quasars: absorption lines}

\section{Introduction}
\setcounter{footnote}{0}

X-ray observations of galaxy clusters and groups have recently revealed a pervasive upper limit on the electron density of the ambient circumgalactic medium (CGM) surrounding a massive galaxy.  Apparently, non-gravitational feedback triggered by radiative cooling and powered by either an active galactic nucleus or supernovae, or maybe a combination of the two, prevents $t_{\rm cool} / t_{\rm ff}$, the ratio of cooling time to freefall time in the ambient medium, from dropping much below $\approx 10$ \citep[e.g.,][]{McCourt+2012MNRAS.419.3319M,Voit_2015Natur.519..203V,Voit+2015ApJ...803L..21V,Hogan_2017_tctff}.  The conventional definition of the cooling time in this critical ratio is $t_{\rm cool} = 3 P / 2 n_e n_i \Lambda$, where $P$ is the gas pressure, $n_e$ and $n_i$ are the electron and ion densities, respectively, and $\Lambda$ is the usual radiative cooling function.  The conventional definition of the freefall time is $t_{\rm ff} = (r / 2g)^{1/2}$, where $g$ is the gravitational acceleration and $r$ is the distance to the bottom of the potential well.  Virtually all galactic systems ranging in mass from $10^{15} \, M_\odot$ down through $10^{13} \, M_\odot$ adhere to this limit \citep{Voit2018_LX-T-R}.

Numerical simulations have shown that the limiting value of $t_{\rm cool} / t_{\rm ff}$ reflects the susceptibility of circumgalactic gas to condensation \citep[e.g.,][]{Sharma_2012MNRAS.420.3174S,Gaspari+2012ApJ...746...94G,Gaspari+2013MNRAS.432.3401G,Li_2015ApJ...811...73L,Prasad_2015ApJ...811..108P}.  In gravitationally stratified media with $t_{\rm cool} / t_{\rm ff} \gg 1$ and a significant entropy gradient, buoyancy suppresses development of a multiphase state \citep{Cowie_1980MNRAS.191..399C}.  Thermal instability does cause small perturbations in specific entropy to grow but results in buoyant oscillations that saturate a fractional amplitude $\sim (t_{\rm cool} / t_{\rm ff})^{-1}$ without progressing to condensation \citep{McCourt+2012MNRAS.419.3319M}.  However, bulk uplift of lower-entropy ambient gas to greater altitudes can induce condensation if it lengthens $t_{\rm ff}$ so that $t_{\rm cool} / t_{\rm ff} \lesssim 1$ {\em within the uplifted gas}.  That condition is relatively easy to satisfy if the global mean ratio is $t_{\rm ff} / t_{\rm cool} \lesssim 10$ but difficult if $t_{\rm ff} / t_{\rm cool} \gtrsim 20$ \citep{Voit_2017_BigPaper}.  Drag can assist condensation by suppressing the damping effects of buoyancy \citep[e.g.,][]{Nulsen_1986MNRAS.221..377N,ps05,McNamara_2016arXiv160404629M}, as can turbulence \citep{Gaspari+2013MNRAS.432.3401G,Voit_2018arXiv180306036V} and magnetic fields \citep{Ji_2018MNRAS.476..852J}.

The implications for massive galaxies are profound.  Feedback from an active galactic nucleus can limit CGM condensation in those systems but requires tight coupling between radiative cooling of the CGM and energy output from the central engine \citep{mn07,McNamaraNulsen2012NJPh...14e5023M}.  A sharp transition to a multiphase state is essential, because it sensitively links the thermal state of the ambient medium on $\sim 10$~kpc scales with feeding of the central black hole on much smaller scales \citep[see][and references therein]{Gaspari_2017MNRAS.466..677G,Voit_2017_BigPaper}.  The feedback loop works like this: If $t_{\rm cool} / t_{\rm ff}$ in the ambient medium is too large, then the black-hole accretion rate is too low for feedback energy to balance radiative cooling.  The specific entropy and cooling time of the ambient medium therefore decline until $t_{\rm cool} / t_{\rm ff}$ becomes small enough for cold clouds to precipitate out of the hot medium.  Those cold clouds then rain down onto the central black hole and fuel a much stronger feedback response that raises $t_{\rm cool}$ in the ambient medium.  Such a system naturally tunes itself to a value of $t_{\rm cool} / t_{\rm ff}$ at which the ambient medium is marginally unstable to precipitation.

This paper proposes some observational tests that can probe whether the precipitation framework for self-regulating feedback also applies to galactic systems in the $10^{11} \, M_\odot$--$10^{13} \, M_\odot$ mass range, in which most of the feedback energy is thought to come from supernovae.  X-ray observations of those systems remain extremely difficult, but the ambient CGM may also leave an imprint on UV absorption-line spectra.  The ions responsible for the O VI and Ne VIII absorption lines observable with Hubble's Cosmic Origins Spectrograph (COS) are not the dominant ones in circumgalactic gas at $\gtrsim 10^6$~K but may still produce detectable signatures.  Consider, for example, the O~VII absorption-line detections of the Milky Way's CGM \citep[e.g.][]{Fang_2006ApJ...644..174F,BregmanLloydDavies_2007ApJ...669..990B,Gupta_2012ApJ...756L...8G,MillerBregman_2013ApJ...770..118M,Fang_2015ApJS..217...21F}, which indicate $N_{\rm OVII} \approx 10^{16} \, {\rm cm^{-2}}$ along lines of sight to extragalactic continuum sources.  Collisional ionization equilibrium at $\sim 10^6 \, {\rm K}$ predicts that $N_{\rm O VI} / N_{\rm O VII} \sim 10^{-2}$  \citep{sd93}.  O~VII absorption-line gas at that temperature would therefore have $N_{\rm O VI} \sim 10^{14} \, {\rm cm^{-2}}$, which is observable with COS.  There may be additional O~VI absorption arising from cooler multiphase gas along those lines of sight, but the ambient gas alone should produce a detectable minimum O~VI signal that depends predictably on the mass of the confining gravitational potential.

It is quite likely that such O~VI absorption lines from the ambient CGM have already been detected.  The most convincing candidates are moderate O~VI lines ($N_{\rm OVI} \sim 10^{14} \, {\rm cm^{-2}}$) associated with broad, shallow Ly$\alpha$ absorption ($N_{\rm HI} \sim 10^{13-14} \, {\rm cm^{-2}}$, $b \sim 100 \, {\rm km \, s^{-1}}$) and comparable Ne~VIII absorption ($N_{\rm NeVIII} \sim 10^{14} \, {\rm cm^{-2}}$).   Such systems sometimes have no associated low-ionization gas \citep[e.g.,][]{Stocke_2013ApJ...763..148S,Werk2016_ApJ...833...54W}. Both the broad Ly$\alpha$ line widths and a collisional-ionization interpretation of the NeVIII/OVI ratios imply gas temperatures $\sim 10^6$~K \citep{Savage2011_OVI_NeVIII_ApJ...743..180S}.  At that temperature, the column densities of the broad Ly$\alpha$ lines imply a total hydrogen column density $N_{\rm H} \sim 10^{20} \, {\rm cm^{-2}}$ \citep{Savage_BroadOVI_2011ApJ...731...14S}.

Section~\ref{sec-CGM_Models} of this paper shows that the precipitation framework, when applied to the Milky Way, predicts that its CGM should indeed have a temperature $\sim 10^6$~K and $N_{\rm H} \sim 10^{20} \, {\rm cm^{-2}}$, nearly independent of projected radius.  The resulting CGM models depend only on the maximum circular velocity of the galaxy's halo, the minimum value of $t_{\rm cool} / t_{\rm ff}$, and surprisingly weakly on heavy-element abundances.  Section~\ref{sec-MilkyWay} presents a detailed comparison of those models with a large variety of Milky-Way data and shows that the models agree with current constraints on the density, temperature, and abundance profiles of the Milky Way's CGM, without any parameter fitting.  In other words, a physically motivated model originally developed to describe feedback regulation of galaxy-cluster cores also aligns with what is currently known about the Milky Way's ambient CGM.  Section~\ref{sec-Columns} then extends that model to predict precipitation-limited O~VI column densities of the ambient CGM in halos ranging in mass from $10^{11} \, M_\odot$ to $10^{13} \, M_\odot$.  For a static CGM, the model gives $N_{\rm OVI} \approx 10^{14} \, {\rm cm^{-2}}$ out to nearly the virial radius across most of the mass range.  However, radial mixing in a dynamic CGM can boost the O~VI column densities to $N_{\rm OVI} \approx 10^{15} \, {\rm cm^{-2}}$ by producing large fluctuations in entropy and temperature that alter the ionization balance.  Section~\ref{sec-SpeculationCirculation} considers the implications of that finding for CGM circulation, supernova feedback, and the dependence of the stellar baryon fraction on halo mass.  Section \ref{sec-Summary} summarizes the paper.

\section{Precipitation-Limited CGM Models}
\label{sec-CGM_Models}

This section presents two simple models for a precipitation-limited CGM.  Both invoke the $t_{\rm cool}/t_{\rm ff} \gtrsim 10$ criterion but make different assumptions about the potential wells and CGM entropy profiles resulting from cosmological structure formation.  The first model was introduced by \citet{Voit2018_LX-T-R}, who used it to calculate $L_X$--$T$ relations for galaxy clusters and groups.   It is extremely simple and serves here to illustrate the basic principles of precipitation-limited models.  The second builds upon the first and is more suitable for predicting absorption-line column densities along lines of sight through the ambient CGM around lower-mass galaxies.

\subsection{The pSIS Model}

The simplest approximation to the structure of a precipitation-limited CGM assumes that the confining potential is a singular isothermal sphere (SIS) characterized by a circular velocity $v_c$ that is constant with radius.  In that case, the corresponding cosmological baryon density profile without radiative cooling or galaxy formation would be
\begin{equation}
  \rho_{\rm cos} (r) = \frac {f_{\rm b} v_c^2} {4 \pi G r^2}
  \; \; ,
\end{equation} 
where $f_{\rm b}$ is the cosmic baryon mass fraction.  Gas with this density profile can remain in hydrostatic equilibrium in the SIS potential if it is at the gravitational temperature $kT_\phi \equiv \mu m_p v_c^2 / 2$, with an entropy profile
\begin{equation}
  K_{\rm SIS} (r) = \frac {\mu m_p} {2} 
  			    \left[ \frac {4 \pi G \mu_e m_p v_c} {f_b} \right]^{2/3}
			    r^{4/3}
  \; \; .
\end{equation}
The slight difference between the $K \propto r^{4/3}$ power-law slope of this approximate cosmological profile and the $K \propto r^{1.1}$ slope found in non-radiative numerical simulations of cosmological structure formation will be addressed in \S \ref{sec-pNFW}.  

As mentioned in the introduction, radiative cooling and the precipitation-regulated feedback that it fuels jointly prevent the ambient cooling time from dropping much below $10 t_{\rm ff}$.  Together, these processes limit the ambient electron density to be no more than about
\begin{equation}
  n_{e,{\rm pre}} (r) = \frac {3 kT} {10 \, \Lambda(T)} \left( \frac {2 n_i} {n} \right) 
  				\frac {v_c} {2^{1/2} r}
  \; \; .
  \label{eq-ne_pre}
\end{equation}
A gas temperature $T = 2 T_\phi$ is required to maintain a gas density profile with $n \propto r^{-1}$ in hydrostatic equilibrium.  Combining these expressions for density and temperature therefore gives a precipitation-limited entropy profile 
\begin{equation}
  K_{\rm pre} (r) = ( 2 \mu m_p)^{1/3} 
  				\left[ \frac {10} {3}
				        \left( \frac {2 n_i} {n} \right)
				         \, \Lambda (2 T_\phi) \right]^{2/3}
				        r^{2/3}
  \label{eq-K_pre}
\end{equation}
that expresses how the minimum specific entropy of the ambient CGM depends on radius.  Notice that equations (\ref{eq-ne_pre}) and (\ref{eq-K_pre}) both assume $\min (t_{\rm cool}/t_{\rm ff}) = 10$, but the limiting $t_{\rm cool}/ t_{\rm ff}$ ratio may also be considered an adjustable parameter of the model.  Observations of galaxy clusters with multiphase gas at their centers show that a large majority of them have $10 \lesssim \min ( t_{\rm cool} / t_{\rm ff} ) \lesssim 20$ \citep{Voit_2015Natur.519..203V,Hogan_2017_tctff}.  Sections \ref{sec-MilkyWay} and \ref{sec-Columns} therefore consider how the predictions of precipitation-limited CGM models change as $\min (t_{\rm cool} / t_{\rm ff})$ shifts through this range.

In the precipitation-limited CGM model originally introduced by \citet{Voit2018_LX-T-R}, which this paper will call the pSIS model, the ambient entropy profile is taken to be the sum of the SIS and precipitation-limited profiles:
\begin{equation}
  K_{\rm pSIS} (r) = K_{\rm SIS} (r) + K_{\rm pre} (r)
  \; \; .
\end{equation}
The assumed temperature profile,
\begin{equation}  
  kT_{\rm pSIS} (r) = \frac {\mu m_p v_c^2 \cdot K_{\rm pSIS} (r)} 
  					{2 K_{\rm SIS}(r) + K_{\rm pre}(r)}
  \; \; ,
\end{equation}
is designed to approach the appropriate limiting values at both small and large radii.  Given these expressions for entropy and temperature, the precipitation-limited electron density profile in the pSIS model is 
\begin{equation}
 n_{e,{\rm pSIS}} (r)  =  \left[ \frac {2 K_{\rm SIS}(r) + K_{\rm pre}(r)} 
 					{\mu m_p v_c^2} \right]^{-3/2}
  \; \; .
\end{equation}  
Multiplying $n_e$ by $2 r_{\rm proj}$ gives the characteristic electron column density along a line of sight through a spherical CGM at a projected radius $r_{\rm proj}$.  This characteristic column density is nearly independent of $r_{\rm proj}$ within the precipitation-limited regions of the pSIS model.

\subsection{The pNFW Model}
\label{sec-pNFW}

Despite its extreme simplicity, the pSIS model makes accurate predictions for the X-ray luminosity-temperature relations among halos in the mass range $10^{12} \, M_\odot$--$10^{15} \, M_\odot$ \citep{Voit2018_LX-T-R}.  However, if one would like to estimate circumgalactic column densities of O~VI and Ne~VIII, the pSIS model has some weaknesses.  Primary among those weakness is its lack of a gas-temperature decline below $\max(T_\phi)$ at large radii.   X-ray observations of galaxy clusters systematically show such a decline \citep[e.g.,][]{Ghirardini_2018arXiv180500042G}, which stems in part from a drop-off in $v_c$ at larger radii and additionally from incomplete thermalization of the kinetic energy being supplied by the incoming accretion flow \citep[e.g.,][]{lkn09}.  There is little direct evidence for a similar outer temperature decline in the ambient gas belonging to halos in the $10^{11} \, M_\odot$--$10^{13} \, M_\odot$ mass range, but if such a decline exists, it can significantly increase the predicted O~VI and Ne~VIII columns, relative to the pSIS model, along any given line of sight through the CGM of a Milky-Way-like galaxy.

Here we construct a slightly less simple alternative, the pNFW model, that addresses those weaknesses.  It assumes a confining gravitational potential with a constant circular velocity at small radii, in order to represent the inner regions of a typical galactic potential well.  At larger radii, the circular-velocity profile declines like that of an NFW halo \citep[e.g.,][]{nfw97} with scale radius $r_s$.  These two circular-velocity profiles are continuously joined at the radius $2.163 r_s$, where the circular velocity of an NFW halo reaches its peak value.  The overall circular-velocity profile is consequently flat at small radii, with $v_c (r) = v_{c,{\rm max}}$ for $r \leq 2.163 r_s$, and declines toward larger radii following
\begin{equation}
  v_c^2(r) = v_{c,{\rm max}}^2 \cdot 4.625 \left[ \frac {\ln (1 + r/r_s)} {r/r_s} - 
  								    \frac {1} {1 + r/r_s} \right] 
  \; \; .
\end{equation}
A halo concentration $r_{200} / r_s = 10$ is assumed, implying that $v_c(r_{200}) = 0.83 \, v_{c,{\rm max}}$, with $r_{200}$ representing the radius encompassing a mean matter density 200 times the cosmological critical density $\rho_{\rm cr}$.  This model gives $r_{200} = (237 \, {\rm kpc}) v_{200}$ and $M_{200} = (1.5 \times 10^{12} \, M_\odot) v_{200}^3$, for $v_{200} \equiv v_{c,{\rm max}} / 200 \, {\rm km \, s^{-1}}$ and $H = 70 \, {\rm km \, s^{-1} \, Mpc^{-1}}$, whereas the gravitational potential in the pSIS model gives $r_{200} = (286 \, {\rm kpc}) v_{200}$ and $M_{200} = (2.6 \times 10^{12} \, M_\odot) v_{200}^3$.  The pSIS and pNFW models are therefore more appropriately compared at similar values of $v_{c,{\rm max}}$ than at similar values of $M_{200}$.

Within this potential well, the baseline entropy profile produced by non-radiative structure formation is taken to be
\begin{equation}
  K_{\rm base} (r) = 1.32 \, \frac {k T_\phi (r_{200})} {\bar{n}_{e,200}^{2/3}} 
				\left( \frac {r} {r_{200}} \right)^{1.1}
				\; \; ,
\end{equation}
where $\bar{n}_{e,200} \equiv 200 f_{\rm b} \rho_{\rm cr} / \mu_e m_p$ is the mean electron density expected within $r_{200}$ \citep{vkb05}.  This expression simplifies to
\begin{equation}
  K_{\rm base} (r) = (39 \, {\rm keV \, cm^2}) \, v_{200}^2
				\left( \frac {r} {r_{200}} \right)^{1.1}
				\; \; ,
\end{equation}
for $v_c(r_{200}) = 0.83 \, v_{c,{\rm max}}$, $H = 70 \, {\rm km \, s^{-1} \, Mpc^{-1}}$,  and $f_{\rm b} = 0.16$.  The modified entropy profile that results from applying the precipitation limit is then
\begin{equation}
  K_{\rm pNFW} (r) = K_{\rm base} (r) + K_{\rm pre} (r)
  \; \; ,
\end{equation} 
with $kT_\phi = \mu m_p v_c^2(r) / 2$ used to determine $\Lambda (2 T_\phi)$ in the calculation of $K_{\rm pre}$ via equation (\ref{eq-K_pre}).

Gas temperature and density in the pNFW model are determined from $K_{\rm pNFW} (r)$ assuming hydrostatic equilibrium.  The integration of $dP/dr$ to find $T(r)$ and $n_e(r)$ depends on a boundary condition that determines the pressure profile.  Choosing $kT(r_{200}) = 0.25 \mu m_p v_{c,{\rm max}}^2$ ensures that the CGM gas temperature drops to roughly half the virial temperature near $r_{200}$, in agreement with observations of the outer temperature profiles of galaxy clusters \citep{Ghirardini_2018arXiv180500042G}.  

\begin{figure*}[t]
\begin{center}
\includegraphics[width=7in, trim = 0.1in 0.1in 0.0in 0.0in]{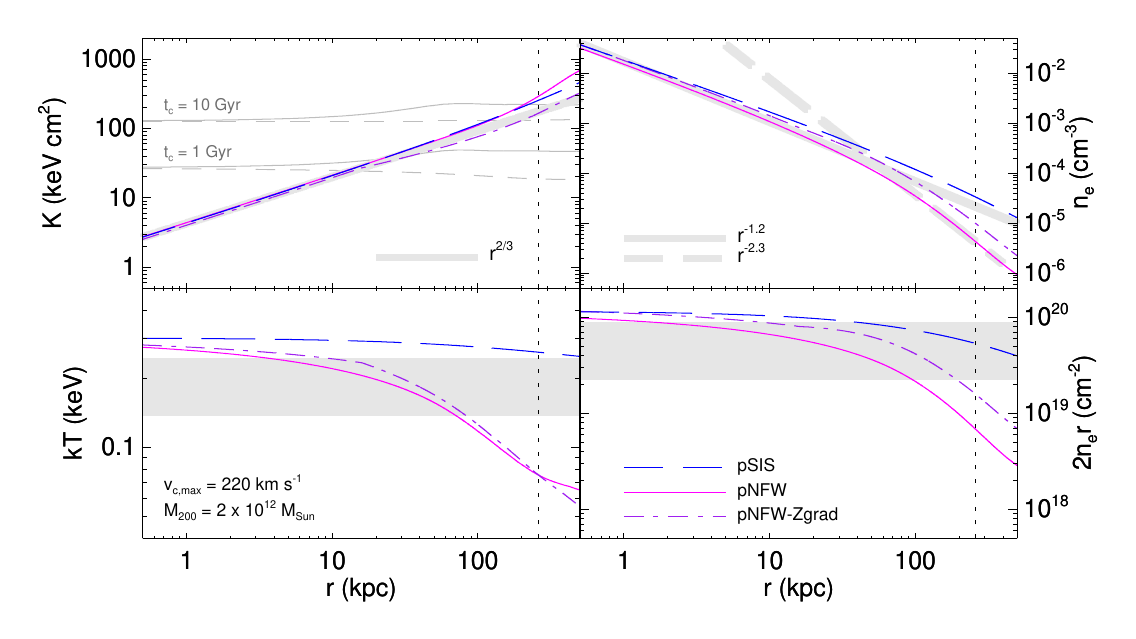} \\
\end{center}
\caption{ \footnotesize 
Comparisons of thermodynamic profiles derived from the pSIS and pNFW models assuming $v_{c,{\rm max}} = 220 \, {\rm km \, s^{-1}}$.  The upper-left panel shows the entropy profiles derived from the pSIS (dashed blue line), pNFW (solid magenta line), and pNFW-Zgrad (dot-dashed purple line) models for the CGM around a Milky-Way-like galaxy with $v_{c,{\rm max}} = 220 \, {\rm km \, s^{-1}}$.  All models have $\min(t_{\rm cool}/t_{\rm ff}) = 10$, and the pSIS and pNFW models have uniform solar abundances.  Thin light grey lines (pSIS: dashed, pNFW: solid) show the corresponding entropy levels at which $t_{\rm cool} = 1$~Gyr and 10~Gyr.  A thick light-grey line with $K \propto r^{2/3}$ illustrates the characteristic power-law slope of a precipitation-limited entropy profile.  The upper-right panel shows electron-density profiles, along with a thick solid line with a slope $\propto r^{-1.2}$ and a dashed solid line with a slope $\propto r^{-2.3}$.  The lower-left panel shows the temperature profiles, with a light-grey band representing showing the two middle quartiles of the Milky Way halo temperature range derived by \citet{HenleyShelton_2013ApJ...773...92H} from X-ray emission observations.  The lower-right panel shows the radial profiles of the characteristic column density $2 n_e(r) r$ associated with ionized circumgalactic gas at radius $r$, along with a light-grey band showing a column-density range derived from X-ray observations of O~VII absorption lines by \citet{MillerBregman_2013ApJ...770..118M} .  A vertical dotted line in each panel shows $r_{200}$ for the pNFW model, and the $M_{200}$ label also refers to that model.
\vspace*{1em}
\label{fig-1}}
\end{figure*}

Figure~\ref{fig-1} compares the radial profiles of $K$, $T$, and $n_e$ predicted by the pSIS and pNFW models for $v_{c,{\rm max}} = 220 \, {\rm km \, s^{-1}}$.  The entropy profiles predicted by the two models are nearly identical, but the pNFW model has a greater temperature gradient, primarily because of the smaller pressure boundary condition applied at $r_{200}$, but also because of the smaller circular velocity at that radius.  Likewise, the density profile of the pNFW model diverges from that of the pSIS model as it approaches $r_{200}$, resulting in a steepening decline of the characteristic column density with radius.  At radii larger than $r_{200}$, the precipitation limit is no longer physically well motivated, because the associated cooling times exceed the age of the universe, as indicated by the thin grey lines in the entropy panel.

\subsection{Assumptions about Abundances}

Inferences of observable CGM properties from the pSIS and pNFW models require supplementary assumptions about the total heavy-element content of the CGM and how it is distributed with radius.  The precipitation framework does not constrain that radial distribution but does make predictions about how the total heavy-element content of the CGM should scale with halo circular velocity.  \citet{Voit_PrecipReg_2015ApJ...808L..30V}  developed models for precipitation-regulated galaxies that link their star-formation rates with enrichment of the CGM.  In those simplistic models, all of the gas associated with a galaxy, including the CGM, is assumed to have a uniform metallicity.  That assumption is what connects the condensation rate of the CGM, and therefore the galactic star-formation rate, to the enrichment of CGM gas.  The resulting stellar mass-metallicity relationship broadly agrees with observations, and so we will adopt that relationship here.  Our fiducial model therefore assumes that a galaxy like the Milky Way has a solar metallicity CGM.

\begin{figure}[t]
\begin{center}
\includegraphics[width=3.5in, trim = 0.1in 0.1in 0.0in 0.0in]{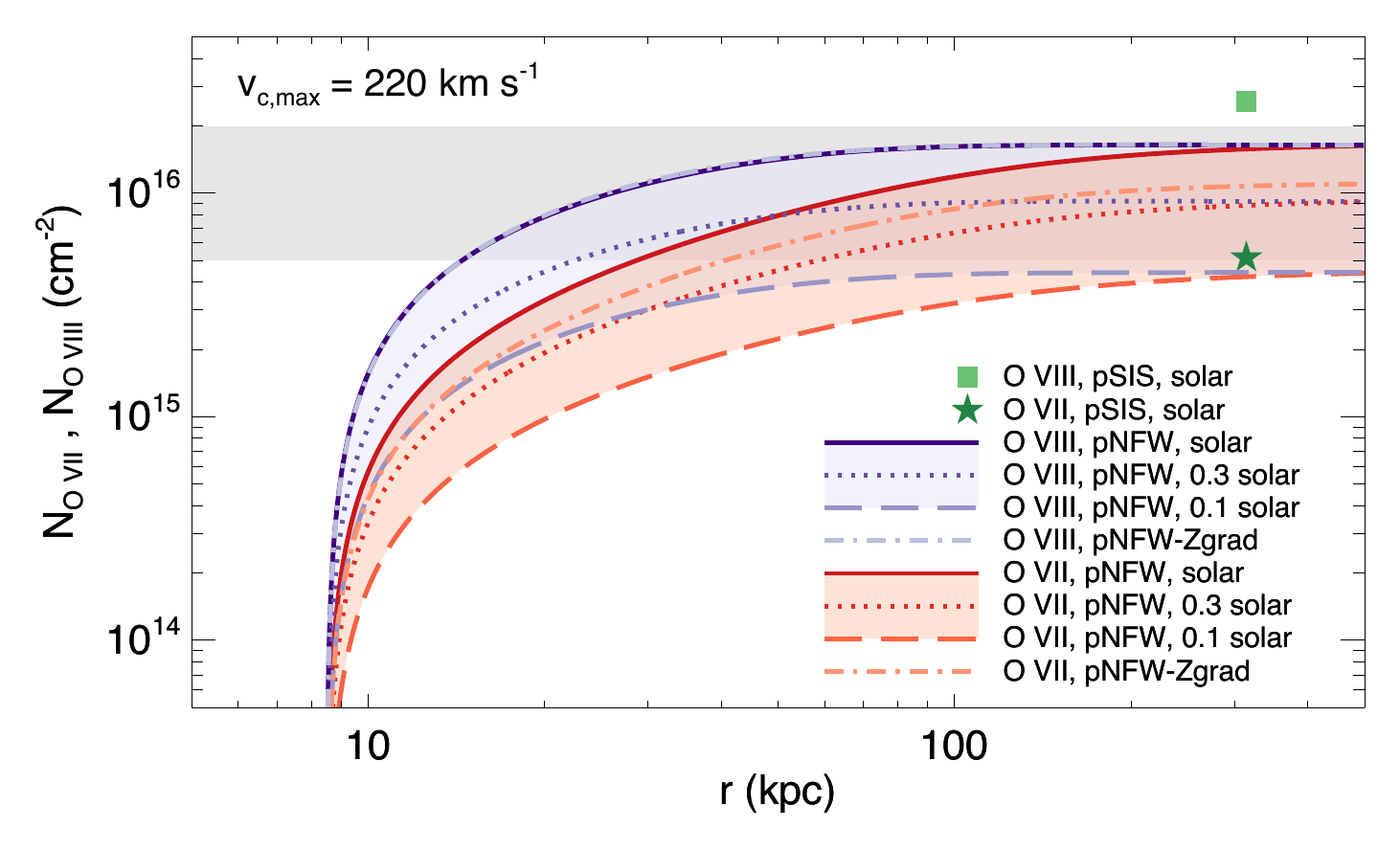} \\
\end{center}
\caption{ \footnotesize 
Comparisons of cumulative O VII and O VIII column densities derived from models with differing abundances along outwardly-directed radial lines of sight starting at 8.5~kpc.  All models assume $v_{c,{\rm max}} = 220 \, {\rm km \, s^{-1}}$ and $\min (t_{\rm cool}/t_{\rm ff}) = 10$.  Red lines and shading show $N_{\rm OVII}$ for pNFW models;  purple lines and shading show $N_{\rm OVIII}$.  Line styles indicate the assumed abundance pattern, including uniform abundances that are solar (solid lines), 0.3 solar (dotted lines), and 0.1 solar (dashed lines), along with the Zgrad abundance gradient (dot-dashed lines).  Green symbols show the predictions for $N_{\rm OVIII}$ (square) and $N_{\rm OVII}$ (star) of pSIS models at $r_{200}$. Grey shading shows the range of Milky-Way $N_{\rm OVII}$ observations compiled by \citet{MillerBregman_2013ApJ...770..118M} .   
\vspace*{3em}
\label{fig-2}}
\end{figure}

However, the predicted absorption-line column densities of highly-ionized elements that emerge from precipitation-limited models are not particularly sensitive to assumptions about the metallicity.  According to equation (\ref{eq-K_pre}), lowering the CGM abundances raises the limiting electron density, and therefore the total CGM column density, by lowering $\Lambda (T)$.  As a result, the predicted column densities of highly-ionized elements have a dependence on abundance that is shallower than linear, as illustrated in Figure~\ref{fig-2}.  The lines in that figure show how the column densities of O~VII and O~VIII predicted by pNFW models rise along lines of sight extending radially outward from a location 8.5~kpc from the center.  Purple lines represent $N_{\rm OVIII}(r)$ and rise more rapidly at smaller radii because of the greater O~VIII fraction there.  Red lines represent $N_{\rm OVII}(r)$ and rise toward $\sim 10^{16} \, {\rm cm^{-2}}$ at larger radii, into the grey shading showing the range of Milky-Way $N_{\rm OVII}$ observations compiled by \citet{MillerBregman_2013ApJ...770..118M}.  Notice that the oxygen column-density predictions of the pNFW models differ by less than a factor of 4, even though the oxygen abundance spans a factor of 10.  Green symbols show the predictions at $r_{200}$ of a solar-abundance pSIS model, in which the CGM temperature exceeds the ambient temperature inferred from X-ray observations and leads to overpredictions of $N_{\rm OVIII}$ and underpredictions of $N_{\rm OVII}$.

Most of the following calculations assume that CGM abundances are independent of radius, but galaxy clusters and groups tend to have declining metallicity gradients, suggesting that CGM metallicity may also depend on radius in less massive galactic systems.  In order to model the $L_X$--$T$ relations of galaxy clusters and groups, \citet{Voit2018_LX-T-R} assumed a metallicity gradient inspired by observations, with $Z(r)/Z_\odot = \min [ 1.0 , 0.3 (r/r_{500})^{-0.5} ]$, where $r_{500}$ is the radius encompassing a mean matter density $500 \rho_{\rm cr}$ and $Z_\odot$ represents solar abundances.  This paper will call a model with that abundance gradient a ``Zgrad" model.

One must also choose a standard ``solar" oxygen abundance.  Values that have been used as standards in recent years range from ${\rm O/H} = 4.6 \times 10^{-4}$ \citep{Asplund_2004A&A...417..751A} through ${\rm O/H} = 8.5 \times 10^{-4}$ \citep{AndersGrevesse_1989GeCoA..53..197A}.  The lower values are in tension with helioseismology, while the higher ones are in tension with 3D solar-atmosphere models \citep[e.g,][]{BasuAntia_2008PhR...457..217B}.  This paper therefore adopts an intermediate value of ${\rm O/H} = 5.4 \times 10^{-4}$ \citep{Caffau_2015A&A...579A..88C} as a standard. 


\section{A Milky-Way Comparison}
\label{sec-MilkyWay}

\begin{figure*}[t]
\begin{center}
\includegraphics[width=7in, trim = 0.1in 0.1in 0.0in 0.0in]{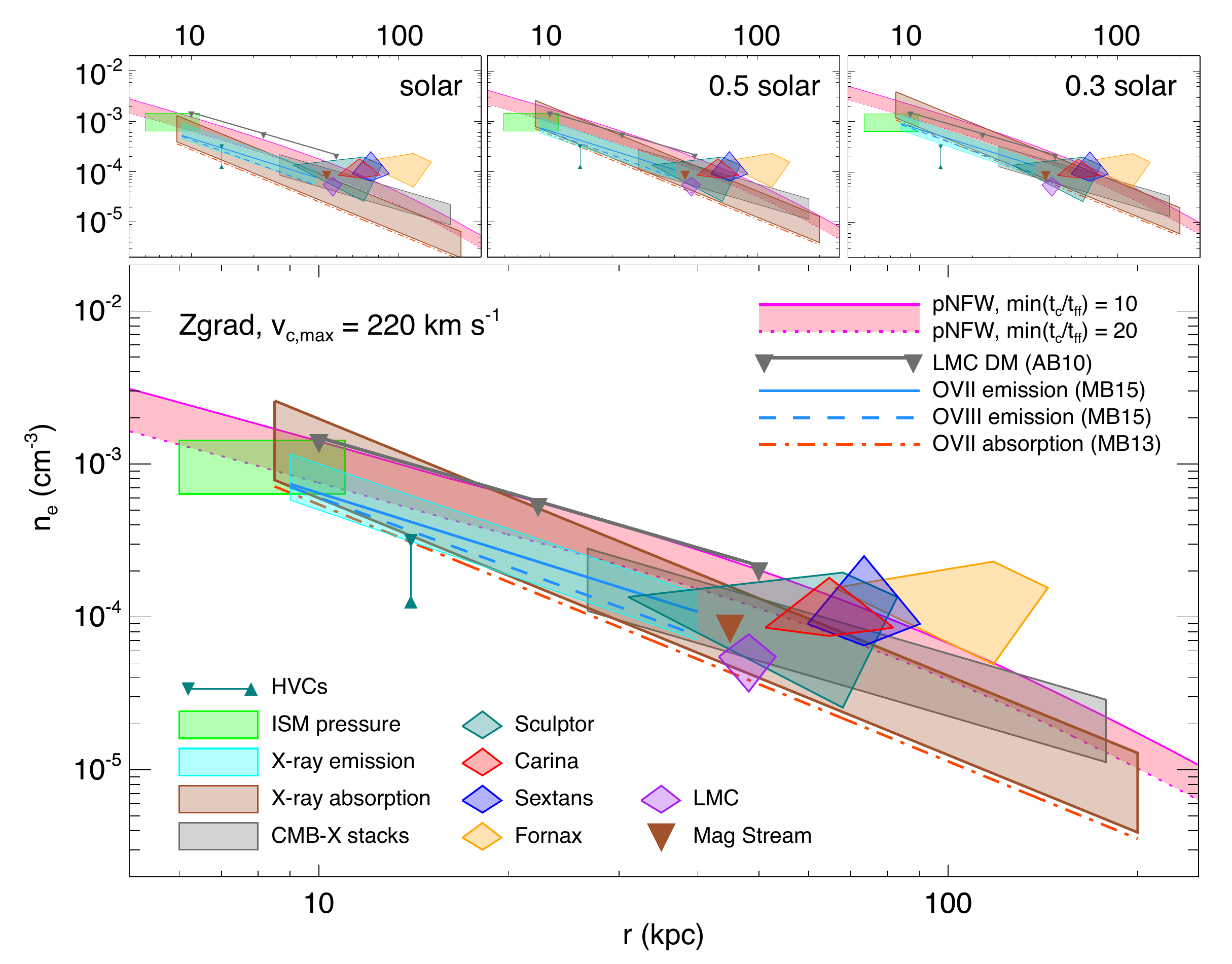} \\
\end{center}
\caption{ \footnotesize 
Comparisons of pNFW models with differing CGM abundances to a broad set of constraints derived mostly from observations of the Milky Way galaxy.  The models assume that a Milky-Way-like potential well with $v_{c,{\rm max}} = 220 \, {\rm km \, s^{-1}}$ confines a hydrostatic CGM with solar abundances (upper left), 0.5 solar abundances (upper middle), 0.3 solar abundances (upper right), and the Zgrad abundance gradient (bottom).  The pink region shows $n_e(r)$ predictions of pNFW models spanning the range $10 \leq \min (t_{\rm cool}/t_{\rm ff}) \leq 20$, with a solid magenta line at  $\min (t_{\rm cool}/t_{\rm ff}) = 10$ and a dotted magenta line at $\min (t_{\rm cool}/t_{\rm ff}) = 20$.  See \S \ref{sec-MilkyWay} for descriptions of the observational constraints.  Note that both the slope and normalization of the model predictions for $n_e(r)$ align remarkably well with the observational constraints, considering that the pNFW predictions are based on physically motivated models of galaxy-cluster cores that have been scaled down to a $2 \times 10^{12} M_\odot$ halo and are not fits to the Milky Way observations. 
\vspace*{1em}
\label{fig-3}}
\end{figure*}

Comparing the precipitation-limited CGM models of \S \ref{sec-CGM_Models} with available data on the Milky Way's ambient CGM reveals a remarkable level of consistency, considering that the precipitation framework was originally developed to describe galaxy clusters and has simply been scaled down to a Milky-Way sized halo.  Figure~\ref{fig-3} shows comparisons of $n_e(r)$ derived from pNFW models based on four different assumptions about the Milky Way's CGM metallicity with a broad set of observational constraints.  The observations generally imply electron density gradients that are similar to the pNFW models, which have $n_e \propto r^{-1.2}$ at small radii and $n_e \propto r^{-2.3}$ at large radii (see Figure \ref{fig-1}).  Differences in assumed abundances affect both the model predictions and most of the observational constraints on $n_e(r)$, but the models are generally most consistent with observations for abundances in the range $0.3 Z_\odot \lesssim Z \lesssim Z_\odot$.  The rest of this section discusses in more detail those observational constraints and how they depend on assumptions about abundances.

\subsection{Interstellar Medium Pressures}

Interstellar thermal gas pressures within a few kpc of the Sun can be robustly measured using ultraviolet observations of absorption lines arising from the three fine-structure levels of the carbon atom's ground state \citep{JenkinsShaya_1979ApJ...231...55J}.  In a comprehensive analysis of the available observational data, \citet{JenkinsTripp_2011ApJ...734...65J} found the mean thermal pressure of the local interstellar medium (ISM) to be $P_{\rm ISM} / k = 3800 \, {\rm K \, cm^{-3}}$, with a dispersion of 0.175 dex and a distribution having wings broader than those expected from a log-normal distribution.  A green rectangle spanning a radial range of 6--10 kpc shows a corresponding range of electron densities derived assuming $n_e = 0.5 \, P_{\rm ISM} / k (2 \times 10^6 \, {\rm K})$.  

The ISM thermal pressure can be considered an upper bound on the CGM thermal pressure at equivalent galactic radii.  While additional forms of ISM support, such as turbulence, magnetic fields, and cosmic-ray pressure, may be comparable to the thermal pressure indicated by the C~I lines, those same sources of additional pressure support are probably at least as important in the CGM.  Given those uncertainties, the pNFW models with $10 \leq \min (t_{\rm cool}/t_{\rm ff}) \leq 20$ agree reasonably well with the ISM pressure constraint, with greater tension arising as the CGM abundances decrease.  However, the pNFW models with $\min (t_{\rm cool}/t_{\rm ff}) \lesssim 20$ and abundances below $0.3 Z_\odot$ imply mean CGM pressures at $\sim 8$~kpc that are significantly greater than the observed ISM pressure.

\subsection{X-ray Emission}

Observations of soft X-ray emission over large portions of the sky consistently indicate that the emissivity-weighted temperature of the Milky Way's hot ambient CGM is approximately $2 \times 10^6$~K \citep[e.g.,][]{KuntzSnowden_2000ApJ...543..195K,McCammon_2002ApJ...576..188M,Gupta_2009ApJ...707..644G,Yoshino_2009PASJ...61..805Y}.  For example, \citet{HenleyShelton_2013ApJ...773...92H}  analyzed {\em XMM-Newton} spectra along 110 lines of sight through the Milky Way and found a fairly uniform median temperature of $2.2 \times 10^6 \, {\rm K}$ with an interquartile range of $0.63 \times 10^6 \, {\rm K}$.  That range is shown with grey shading in the lower-left panel of Figure~\ref{fig-1}.  It is consistent with the Milky Way CGM temperature predicted by the pNFW model for radii from 3~kpc to 70~kpc but is inconsistent with the pSIS model, which predicts hotter temperatures.  

\citet{HenleyShelton_2013ApJ...773...92H}  also found a spread in emission measure ranging over $\sim (0.4$--$7) \times 10^{-3} \, {\rm cm^{-6} \, pc}$, with a median of $1.9 \times 10^{-3} \, {\rm cm^{-6} \, pc}$, assuming solar abundances.  Emission measure generally increases toward the center of the galaxy but is not strongly dependent on galactic latitude, indicating that the gas distribution is more spherical than disk-like.

\citet{MillerBregman_2015ApJ...800...14M}  used an even larger sample of O~VII and O~VIII emission-line observations compiled by \citet{HenleyShelton_2012ApJS..202...14H} to constrain the radial density distribution of the line-emitting gas.  They selected a subset of 649 {\em XMM-Newton} spectra sampling the entire sky and fit them with a model assuming constant-temperature gas at $T = 10^{6.3} \, {\rm K}$ and a power-law density profile $n_e \propto r^{-3 \beta}$.  This isothermal power-law model yielded an excellent fit to the O~VIII emission for $\beta = 0.50 \pm 0.03$, assuming optically-thin emission, and $\beta = 0.54 \pm 0.03$ after accounting for potential optical-depth effects.  

Dashed blue lines in Figure~\ref{fig-3} show the best fit from \citet{MillerBregman_2015ApJ...800...14M}  to optically-thin O~VIII emission from a solar-abundance plasma.  In the panels corresponding to $0.5 Z_\odot$ and $0.3 Z_\odot$, the density normalization of that fit has been multiplied by $(Z/Z_\odot)^{-1/2}$, because the line intensity scales $\propto n_e^2 (Z/Z_\odot)$.  In the panel showing the Zgrad model, the abundance correction corresponds to a uniform abundance of $0.5 Z_\odot$.  Each of the lines representing emission constraints extends from 9~kpc to 40~kpc because integrating over larger radii increases the emission measure by $< 10$\%.  Notice that the best-fitting power law from \citet{MillerBregman_2015ApJ...800...14M}  has a slope ($n_e \propto r^{-1.5}$) that is similar to the pNFW models within that radial range but has a slightly lower normalization.

Solid blue lines in Figure~\ref{fig-3} show abundance-corrected versions of the best fit by \citet{MillerBregman_2015ApJ...800...14M}  to optically-thin O~VII emission, which has $\beta = 0.43 \pm 0.01$.  Each of those lines therefore illustrates a density profile with $n_e \propto r^{-1.29}$, which is also quite similar to the electron-density profile shape predicted by the pNFW models in the 9~kpc to 40~kpc region. \citet{MillerBregman_2015ApJ...800...14M}  report that their best fit to the O~VII emission data is poorer than their best fit to the O~VIII emission.  In order to obtain acceptable $\chi^2$ values, they had to add systematic scatter of roughly a factor of 2 to their error budget, suggesting that the gas responsible for much of the O~VII emission is inhomogeneous.  Relatively modest variations in gas temperature within the observed temperature range can potentially produce that inhomogeneity, because the O~VII ionization fraction rises by more than a factor of 6 as the CGM temperature declines through the \citet{HenleyShelton_2013ApJ...773...92H}  range from $2.83 \times 10^6$~K to $1.57 \times 10^6$~K.  Over the same temperature range, the O~VIII ionization fraction changes by less than a factor of 2.

\citet{MillerBregman_2015ApJ...800...14M}  hypothesized that the difference in power-law slope between the O~VIII and O~VII best fits may arise from a temperature gradient, because of how the ionization fractions change with temperature.  Their Figure~13 shows that the necessary temperature gradient is approximately $T \propto r^{-0.08}$ if the temperature at 8.5~kpc is held fixed at $2 \times 10^6 \, {\rm K}$.  In that same vicinity, the pNFW models have a temperature slope similar to $T \propto r^{-0.13}$, with a temperature $\approx 2.6 \times 10^6 \, {\rm K}$ at 8.5~kpc.

The broad cyan strips in Figure~\ref{fig-3} are based on the range of emission-measure observations found by \citet{HenleyShelton_2013ApJ...773...92H}  and have a power-law slope $n_e \propto r^{-1.4}$, in between the slopes derived from Miller \& Bregman's O~VII and O~VIII best fits.  Each cyan strip shows a range of electron density profiles corresponding to an emission-measure range $(1$--$4) \times 10^{-3} \, {\rm cm^{-6} \, pc}$, multiplied by a metallicity correction factor $[\Lambda(T,Z_\odot))/\Lambda(T, Z)]^{1/2}$ with $T = 2.2 \times 10^6 \, {\rm K}$.  For all of the assumed metallicities, the high end of this emission-measure range is generally more consistent with the pNFW models than the low end.  In that context, it is worth noting that the median emission measure from \citet{HenleyShelton_2013ApJ...773...92H}  falls slightly below the emission measures found by some other studies \citep[e.g.,][]{Yoshino_2009PASJ...61..805Y,Gupta_2009ApJ...707..644G}.

\subsection{X-ray Absorption}

X-ray observations of O~VII and O~VIII absorption lines provide complementary constraints on the electron-density profile that help to break model degeneracies.  \citet{MillerBregman_2013ApJ...770..118M}  undertook a comprehensive analysis of the available O~VII absorption data, which they extended in \citet{MillerBregman_2015ApJ...800...14M}.  Assuming optically-thin absorption and a power-law density profile with a constant $n_{\rm OVII}/n_e$ ratio, they found a best-fit density profile with $\beta = 0.56_{-0.12}^{+0.10}$.  When attempting to correct for saturation assuming an absorption-profile velocity width $b = 150 \, {\rm km \, s^{-1}}$, they found $\beta = 0.71_{-0.14}^{+0.13}$ for the whole data set and $\beta = 0.60_{-0.13}^{+0.12}$ using only the observations with signal-to-noise $> 1.1$.  The lines of sight along directions that pass within $< 8.5$~kpc of the galactic center tend to receive the greatest saturation corrections, while the saturation corrections along lines of sight pointing away from the center tend be small.  The $n_e \propto r^{-1.68}$ power-law density profile found without saturation correction may therefore be more representative of radii $> 8.5$~kpc, and this paper will adopt it for comparisons with the pNFW models.

Dot-dashed (orange-red) lines in Figure~\ref{fig-3} show the \citet{MillerBregman_2013ApJ...770..118M}  electron density profiles derived from O~VII absorption assuming no saturation.  The abundance corrections are $\propto Z^{-1}$, with a correction for a uniform abundance of $0.5 Z_\odot$ applied in the Zgrad panel.  Those power-law profiles have slopes similar to the pNFW models in the 8.5--200~kpc range and lie significantly below the pNFW predictions.  However, the constant O~VII ionization fraction of 0.5 assumed by \citet{MillerBregman_2013ApJ...770..118M}  is inconsistent with the pNFW models, in which collisional ionization equilibrium gives an ionization fraction $n_{\rm OVII}/n_{\rm O} \lesssim 0.2$ at $< 10$~kpc and $n_{\rm OVII}/n_{\rm O} \gtrsim 0.4$ at $\sim 40$~kpc.  The tendency for the O~VII ionization fraction in the pNFW models to be $< 0.5$ at radii $\lesssim 30$~kpc can also be seen in Figure~\ref{fig-2}, which shows that the cumulative O~VIII column density along directions away from the galactic center rises more sharply with radius than the cumulative O~VII column density in the 8.5~kpc to $\gtrsim 30$~kpc interval.

A proper comparison of the \citet{MillerBregman_2013ApJ...770..118M} data set with the pNFW models therefore requires an upward renormalization of the electron-density profiles derived from them.  Brown strips in Figure~\ref{fig-3} show electron-density profiles with normalizations determined assuming collisional ionization equilibrium at the temperatures given by the pNFW models.  All of the brown strips share the same power-law slope ($n_e \propto r^{-1.68}$) as the best fitting optically-thin model from \citet{MillerBregman_2013ApJ...770..118M}.  The lower edge of each brown strip is normalized so that $N_{\rm OVII} = 5 \times 10^{15} \, {\rm cm^{-2}}$, corresponding to the low end of the \citet{MillerBregman_2013ApJ...770..118M}  data set.   The upper edge of each strip is normalized so that $N_{\rm OVII} = 1.5 \times 10^{16} \, {\rm cm^{-2}}$, which is the weighted mean column density found by \citet{Gupta_2012ApJ...756L...8G}, after they corrected for saturation.  These brown strips generally agree well with the pNFW model in both normalization and slope, with slightly better agreement for sub-solar CGM metallicities.

\citet{Gupta_2012ApJ...756L...8G} also presented O~VIII equivalent-width measurements, showing that they are comparable to the O~VII equivalent widths.  This finding is consistent with the pNFW models shown in Figure~\ref{fig-2}, even though the ratio of O~VII  to O~VIII is not constant with radius.  More recently, Nevalainen et al. (2017) have published {\em XMM-Newton} absorption-line observations of O~IV, O~V, O~VII, and O~VIII along a particularly well-observed line of sight toward PKS 2155-304.  They derived independent O~VII and O~VIII column-density measurements from the four different {\em XMM-Newton} detectors, finding column densities within a factor of two of $1 \times 10^{16} \, {\rm cm^{-2}}$ for both lines, again consistent with the pNFW models in Figure~\ref{fig-2} for CGM abundances in the $(0.3$--$1)~Z_\odot$ range.

\subsection{LMC Dispersion Measure}

Observations of the dispersion measure toward pulsars in the Large Magellanic Cloud place an upper limit on the normalization of the $n_e$ profile within 50~kpc of the galactic center.  \citet{AndersonBregman_2010ApJ...714..320A} found the dispersion measure attributable to the CGM in that radial interval to be no greater than $2.3 \times 10^{-2} \, {\rm cm^{-3} \, kpc}$.  Grey lines with inverted triangles illustrate this upper limit in the four panels of Figure~\ref{fig-3}, assuming a density profile with $n_e \propto r^{-1.2}$, as in the inner parts of the pNFW models.

\citet{MillerBregman_2013ApJ...770..118M,MillerBregman_2015ApJ...800...14M} showed that this upper limit, which does not depend on metallicity, places interesting constraints on the CGM metallicity when combined with inferences of $n_e(r)$ from their O~VII and O~VIII data sets.   They found that a CGM metallicity $\gtrsim 0.3 Z_\odot$ was necessary to satisfy all of their constraints.  Likewise, the LMC dispersion-measure limits place interesting constraints on the allowed metallicities of pNFW models for the Milky Way's CGM.  The entire $10 \leq \min ( t_{\rm cool} / t_{\rm ff} ) \leq 20$ range of solar-metallicity models can satisfy the dispersion-measure constraint, but the models come into increasing tension with the constraint as metallicity decreases.  In the $0.3 Z_\odot$ case, only pNFW models with $\min (  t_{\rm cool} / t_{\rm ff}  ) \gtrsim 20$ are permitted.

\subsection{Ram-Pressure Stripping}

Additional metallicity-independent constraints come from ram-pressure stripping models of dwarf galaxies that orbit the Milky Way.  Figure~\ref{fig-3} uses diamond-like polygons to illustrate those constraints.  The horizontal span of each polygon shows the uncertainty in radius of the orbital pericenter; the vertical span shows the uncertainty in inferred CGM density at the pericenter.  A purple polygon shows constraints derived from the LMC by \citet{Salem_2015ApJ...815...77S}.  Red and blue polygons show constraints derived by \citet{Gatto_2013MNRAS.433.2749G} from the Carina and Sextans dwarf galaxies, respectively.  Orange and blue polygons show constraints derived by \citet{Grcevich_2009ApJ...696..385G} from the Fornax and Sculptor dwarf galaxies, respectively.  Constraints based on the other two dwarf galaxies modeled by \citet{Grcevich_2009ApJ...696..385G} are not shown because they are too weak to be interesting in this context.  As a group, these ram-pressure constraints tend to be in tension with the uncorrected density profiles inferred from the O~VII and O~VIII data by \citet{MillerBregman_2013ApJ...770..118M,MillerBregman_2015ApJ...800...14M}.  They are in better agreement with the pNFW models, particularly at the lower end of the CGM metallicity range allowed by the dispersion-measure constraints.  The model with a metallicity gradient (pNFW-Zgrad) is the most successful at satisfying both the dispersion-measure and ram-pressure constraints.

\subsection{High-Velocity Clouds}

Circumgalactic pressures can be derived from 21~cm observations of H~I in high-velocity clouds with the help of assumptions about their distance and shape.  If the clouds are roughly spherical, their extent along the line of sight can be estimated from their transverse size, given a distance estimate.  A column-density measurement can then be converted into a density measurement, which becomes a pressure measurement when combined with information about the cloud's temperature.  The pressures inferred by \citet{Putman_2012ARA&A..50..491P} from such observations of high-velocity clouds at distances $\sim 10$--15~kpc from the galactic center are $10^{2.7} \, {\rm K \, cm^{-3}} \lesssim P/k \lesssim 10^{3.1} \, {\rm K \, cm^{-3}}$.  In Figure~\ref{fig-3}, teal line segments bounded by triangles show the CGM density constraints that result from assuming that those clouds are in pressure equilibrium with an ambient medium at $2.2 \times 10^6 \, {\rm K}$.  They tend to indicate ambient densities lower than those derived from other constraints, with increasing tension as the assumed CGM metallicity declines.

\subsection{Magellanic Stream}

Similar constraints on ambient pressure can be derived from 21~cm observations of clouds in the Magellanic Stream.  Inverted brown triangles in Figure~\ref{fig-3} show ambient density constraints that follow from pressure estimates by \citet{Stanimirovich_2002ApJ...576..773S}, who considered them upper limits on the actual thermal pressure because other forms of pressure, such as ram pressure, could also be contributing to cloud compression.  Their electron-density constraint at an assumed distance of 45~kpc, which has been adjusted here for consistency with the $1.8 \times 10^6$~K ambient temperature predicted at that distance by pNFW models, is similar to the ambient densities inferred from ram-pressure stripping of dwarf galaxies.

\subsection{CMB/X-ray Stacking}

The final set of constraints shown in Figure~\ref{fig-3} is derived from galaxies more massive than the Milky Way.  \citet{Singh_stacks_2018MNRAS.478.2909S} combined stacked X-ray observations of galaxies with halo masses in the $10^{12.6} M_\odot$--$10^{13.0} M_\odot$ range from \citet{Anderson_2015MNRAS.449.3806A} and stacked CMB observations from {\em Planck} in that same mass range \citep{Planck_LRGstacks_2013A&A...557A..52P}.  By jointly fitting those data sets with simple CGM scaling laws, \citet{Singh_stacks_2018MNRAS.478.2909S}  found a best-fit density slope $n_e \propto r^ {-1.2}$ and a best-fitting CGM temperature scaling law that extrapolates to $\approx 2.2 \times 10^6 \, {\rm K}$ at the mass scale of the Milky Way.

Grey strips in Figure~\ref{fig-3} show where the best power-law fits of Singh et al. (2018) to CGM density profiles fall when extrapolated to a pNFW model with $M_{200} = 2 \times 10^{12} \, M_\odot$ and $v_{c,{\rm max}} = 220 \, {\rm km \, s^{-1}}$.  Metallicity corrections have been made because the original power-law fits assumed a metallicity of $0.2 Z_\odot$.  They have therefore been multiplied by $[\Lambda(T,0.2 Z_\odot))/\Lambda(T, Z)]^{1/2}$, with $T = 2.2 \times 10^6 \, {\rm K}$, to account for the effects of metallicity on X-ray emission.  The strips span the radial range $(0.15$--$1) r_{500}$ because they are derived from projected data excluding the core region at $< 0.15 r_{500}$.  The vertical span of each strip reflects an uncertainty range extending a factor of 1.6 in each direction, corresponding to the uncertainty range of the CGM baryonic gas fraction in the fits of \citet{Singh_stacks_2018MNRAS.478.2909S}.

\subsection{Comparison Summary}

\begin{figure}[t]
\begin{center}
\includegraphics[width=3.5in, trim = 0.1in 0.1in 0.0in 0.0in]{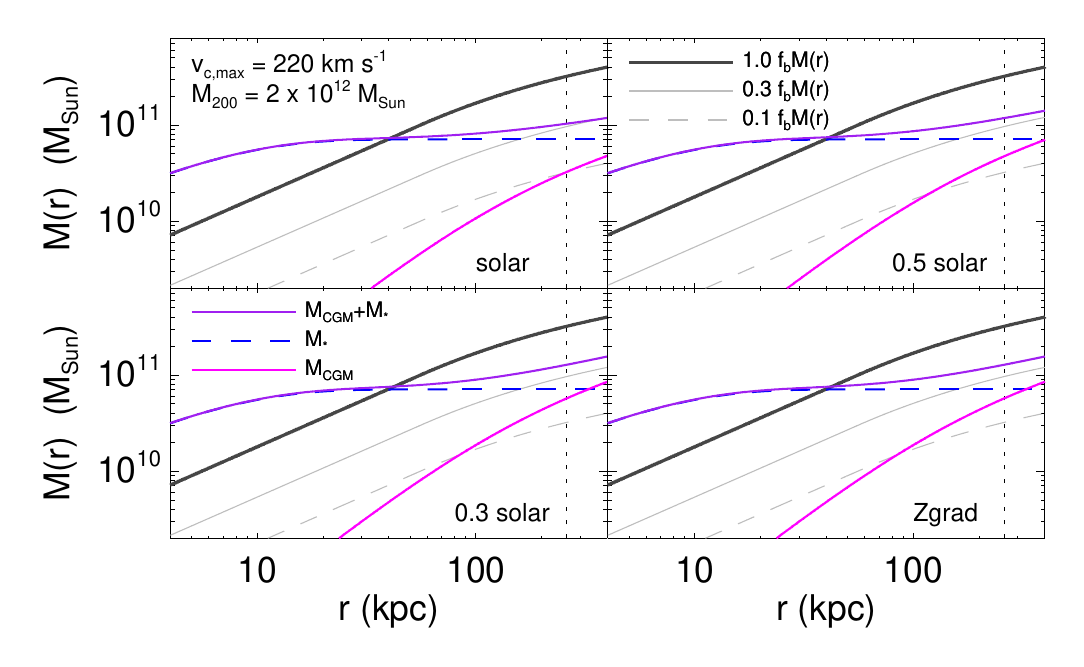} \\
\end{center}
\caption{ \footnotesize 
Baryonic mass profiles of pNFW models with different abundance patterns.  Each model has $v_{\rm c,max} = 220 \, {\rm km \, s^{-1}}$, $M_{200} = 2 \times 10^{12} \, M_\odot$, and $\min (t_{\rm cool} / t_{\rm ff}) = 10$.  Magenta lines show the cumulative mass of ambient CGM ($M_{\rm CGM}$) within $r$.  Dashed blue lines show stellar mass profiles assuming that $M_*$ is consistent with $v_{\rm c,max} = 220 \, {\rm km \, s^{-1}}$ at small radii and asymptotically approaches $7 \times 10^{10} \, M_\odot$ at large radii.   Purple lines show total baryonic mass profiles equal to the sum of $M_{\rm CGM}$ and $M_*$.  Thick charcoal lines show total mass profiles multiplied by the cosmic baryon fraction $f_{\rm b}$, solid light grey lines show $0.3 f_{\rm b} M(r)$, and dashed light grey lines show $0.1 f_{\rm b} M(r)$.  Shifting to $\min (t_{\rm cool} / t_{\rm ff}) = 20$ would move the $M_{\rm CGM}$ profiles down by a factor $\approx 2$.  Note that feedback must push at least half of the halo's baryons outside of $r_{200}$ (indicated by the dotted black line).  However, the gas mass of the interstellar medium and the low-ionization phases of the CGM is not accounted for in these plots.  
\vspace*{1em}
\label{fig-4}}
\end{figure}

Taken as a whole, these comparisons of observations with the pNFW models show that the CGM of the Milky Way is plausibly precipitation-limited, in a manner similar to the multiphase cores of galaxy clusters and central group galaxies, which also tend to have $10 \lesssim \min ( t_{\rm cool} / t_{\rm ff} ) \lesssim 20$.  There are some points of tension with the data that need to be better understood through attempts to fit those data sets with parametric pNFW models.  However, we will leave that task for the future.  The main objective of this section has been to validate the pNFW models through comparisons with Milky Way before relying on them to make predictions for UV absorption lines from the CGM of galaxies with halo masses $10^{11} \, M_\odot$--$10^{13} \, M_\odot$. 

For reference, Figure~4 shows how the total baryonic mass enclosed within a given radius rises toward radii $> r_{200}$.  The stellar mass of this Milky-Way-like galaxy is assumed to be $7 \times 10^{10} \, M_\odot$, with a mass distribution giving $v_c = 220 \, {\rm km \, s^{-1}}$ at small radii.  Gas-mass profiles ($M_{\rm CGM}$) in the figure are derived from pNFW models assuming $\min ( t_{\rm cool} / t_{\rm ff} ) = 10$.  The total baryonic mass within $r_{200}$ predicted by the solar-abundance pNFW model corresponds to $\sim 30$\% of the cosmic baryon fraction and rises to $\sim 40$\% in the pNFW-Zgrad model.  Potential contributions from the galactic ISM and lower-ionization phases of the CGM are not included in these estimates but are unlikely to close the baryon budget.  Therefore, a galaxy like the Milky Way must push at least 50\% of its baryons beyond $r_{200}$ in order to satisfy the precipitation limit.

\newpage

\subsection{Relationships to Similar Models}

\begin{figure}[t]
\begin{center}
\includegraphics[width=3.5in, trim = 0.1in 0.1in 0.0in 0.0in]{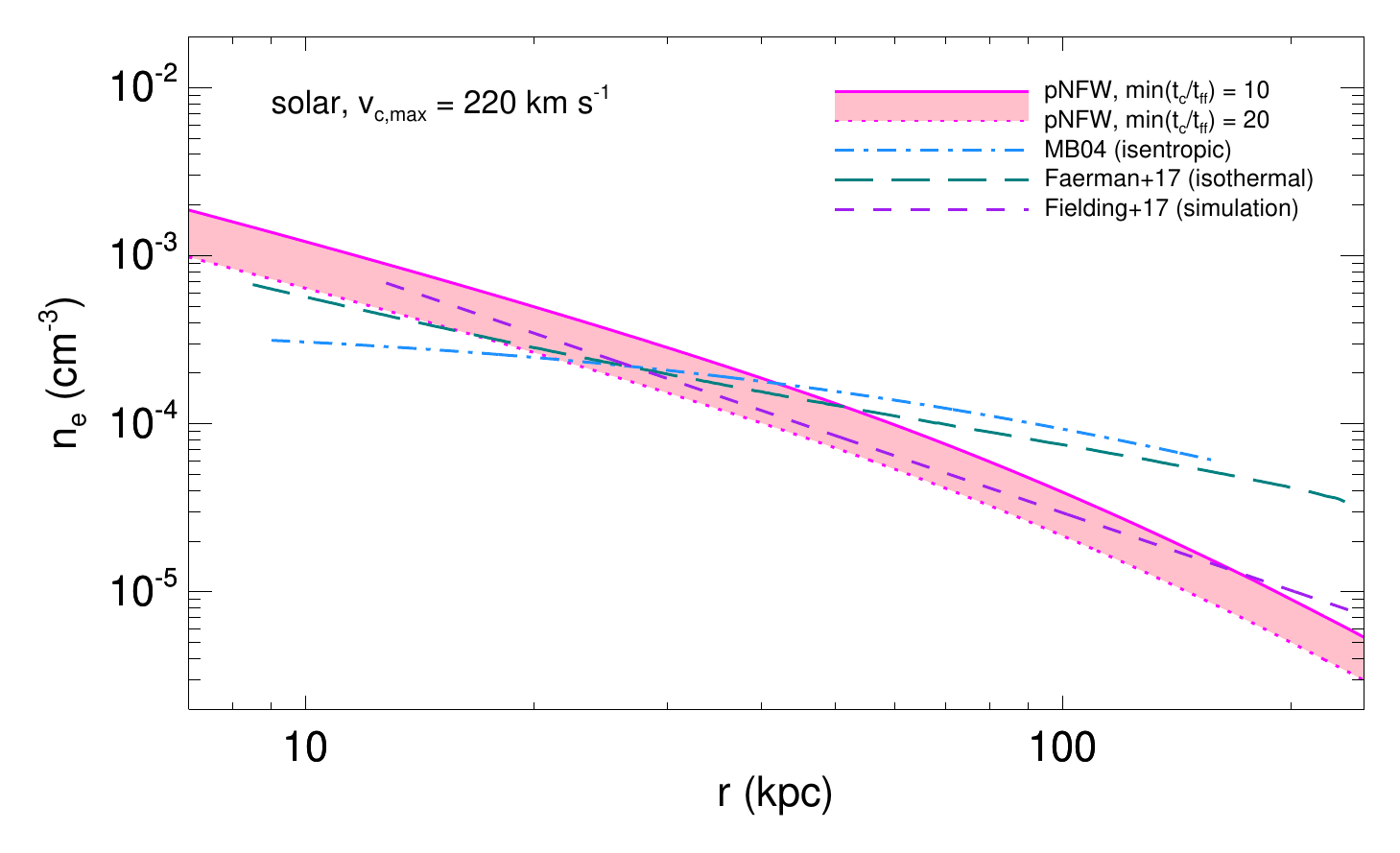} \\
\end{center}
\caption{ \footnotesize 
Comparison of models for the Milky Way's CGM.   Magenta lines and the pink band represent the same solar-metallicity pNFW models shown in the upper-left panel of Figure~\ref{fig-3}.   A blue dot-dashed line shows the adiabatic CGM model from \citet{MallerBullock_2004MNRAS.355..694M}, for their default metallicity of $0.1 Z_\odot$.  A teal long-dashed line shows the isothermal CGM model from \citet{Faerman_2017ApJ...835...52F}, which has a temperature of $1.5 \times 10^6$~K.  A purple short-dashed line shows the ambient electron density from the idealized supernova-feedback simulation of \citet{Fielding_2017MNRAS.466.3810F}, which held the CGM metallicity constant at $Z = Z_\odot / 3$.
\vspace*{1em}
\label{fig-4A}}
\end{figure}

Other models for the Milky Way's CGM based on different assumptions have made comparable predictions.  For example, the model of \citet{Faerman_2017ApJ...835...52F} assumes that the CGM is isothermal at $1.5 \times 10^6$~K with 60--$80 \, {\rm km \, s^{-1}}$ of turbulence and log-normal temperature fluctuations with a dispersion $\sigma_{\ln T} = 0.3$.  Figure~\ref{fig-4A} shows that this isothermal model is similar to the pNFW model at 20--60~kpc but has a flatter electron-density profile and a larger CGM mass inside of $r_{200}$.   Likewise, the isentropic CGM model of \citet{MallerBullock_2004MNRAS.355..694M} also has a flatter profile than the pNFW model and a greater CGM mass within $r_{200}$.  Both of those other models are in considerable tension with the electron-density profiles inferred from X-ray spectroscopy by \citet{MillerBregman_2013ApJ...770..118M,MillerBregman_2015ApJ...800...14M}.   In contrast, the idealized Milky-Way galaxy simulated by \citet{Fielding_2017MNRAS.466.3810F}, in which supernova-driven winds regulate the structure of the CGM, has an ambient density profile ($n_e \appropto r^{-1.5}$) consistent with the profile slopes derived from both X-ray spectroscopy and precipitation-limited models.  

\section{Ambient O~VI Column Densities}
\label{sec-Columns}

The preceding section demonstrated that precipitation-limited models for the Milky Way's ambient CGM are compatible with the available observational constraints.  This section uses those models to make predictions for the column densities of O~VI, Ne~VIII, and N~V in the ambient CGM around galaxies in halos ranging from $10^{11} \, M_\odot$--$10^{13} \, M_\odot$, so that the precipitation framework can be tested with UV absorption-line observations.  It first considers a static CGM with gas temperatures and ionization states that are uniform at each radius.  Under those conditions, the models predict that the ambient CGM has $N_{\rm OVI} \approx 10^{14} \, {\rm cm^{-2}}$ over wide ranges in projected radius, halo mass, and CGM metallicity.  However, the observed velocity structure of the O~VI lines clearly shows that the CGM is not static.  Gas motions in the CGM can produce temperature fluctuations that broaden the range of ionization states expected at each radius.  This section shows that accounting for temperature fluctuations leads to O~VI predictions that can rise as high as $N_{\rm OVI} \approx 10^{15} \, {\rm cm^{-2}}$ in $10^{12} \, M_\odot$ halos and may offer opportunities to probe how disturbances propagating through the CGM stimulate condensation and production of lower-ionization gas. 

\subsection{Static CGM}
\label{sec-Static}

\begin{figure*}[t]
\begin{center}
\includegraphics[width=7.0in, trim = 0.1in 0.1in 0.0in 0.0in]{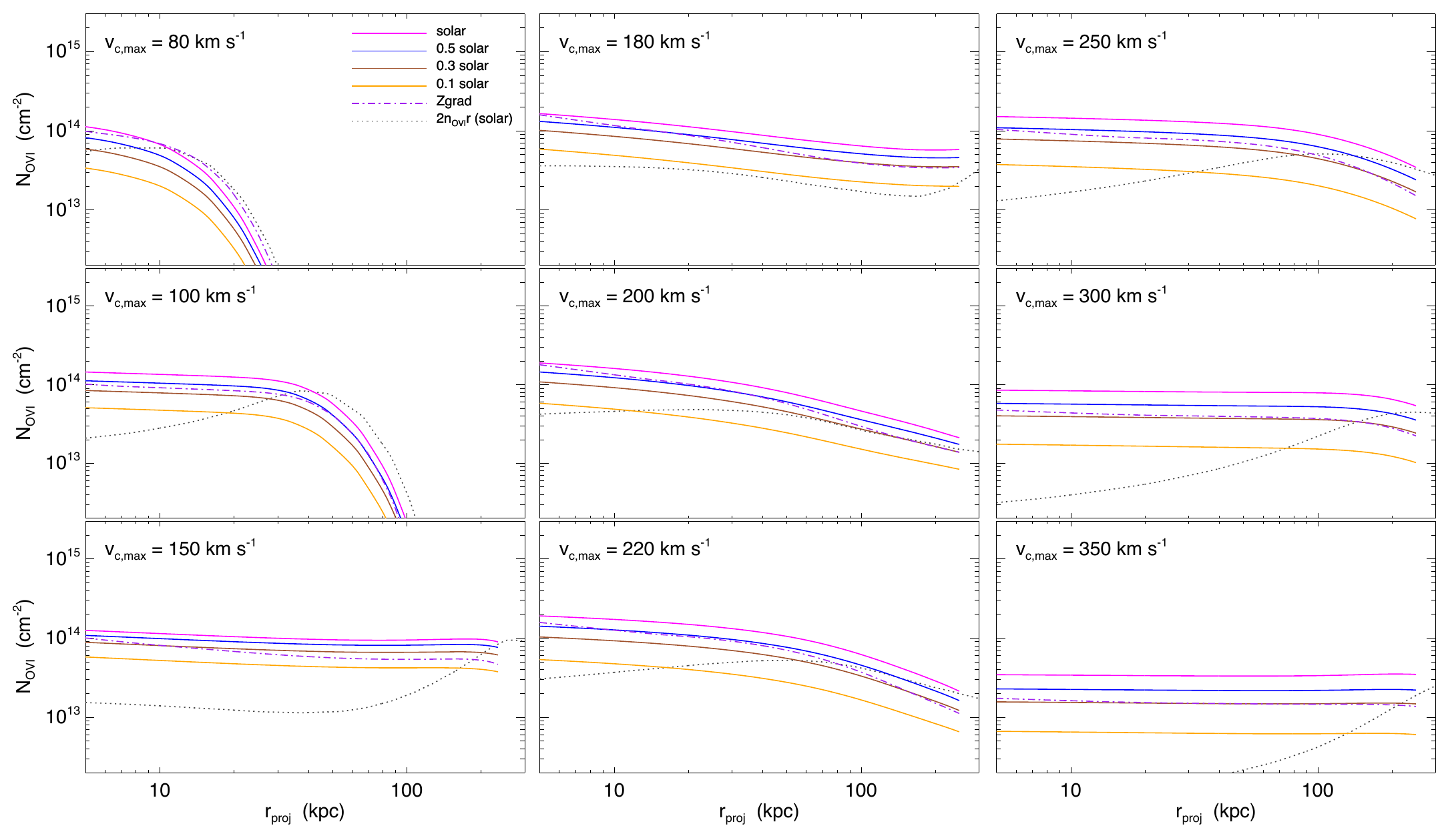} \\
\end{center}
\caption{ \footnotesize 
Profiles of ambient O~VI column density as a function of impact parameter $r_{\rm proj}$ predicted by pNFW models for a static CGM.  Each model has $\min (t_{\rm cool} / t_{\rm ff}) = 10$ and a maximum circular velocity shown by the label.  Solid lines show models with abundances of $Z_\odot$ (magenta), $0.5 Z_\odot$ (blue), $0.3 Z_\odot$ (brown), and $0.1 Z_\odot$ (orange).  Dot-dashed purple lines show models with the Zgrad abundance gradient.  Dotted black lines showing the characteristic column density $2 n_{\rm OVI} r$ reflect the proportional contribution of each spherical shell to the total column density.  Generally, the projected hydrogen column density gradually decreases with $r_{\rm proj}$, but $2 n_{\rm OVI} r$ sometimes increases with $r_{\rm proj}$ because of how the decline in ambient temperature with radius affects the O~VI ionization fraction.
\vspace*{1em}
\label{fig-5}}
\end{figure*}

The CGM models presented in \S \ref{sec-CGM_Models} are completely hydrostatic, and so have a unique temperature at each radius.  In collisional ionization equilibrium, that temperature determines the ion fractions at each radius.  Integration along a CGM line of sight at a particular projected radius $r_{\rm proj}$ to find the column density of each ion is then straightforward but requires some assumptions about the limits of integration.  The column density predictions presented here apply two limits.  First, the spherical CGM models to be integrated are truncated at $2 r_{100}$, where $r_{100}$ contains a mean mass density $100 \rho_{\rm cr}$.  This choice ensures that the line-of-sight integration does not extend far beyond the virialized region around the galaxy, outside of which the pNFW models are unlikely to be valid.  Second, the integration is limited to within a physical radius of 500 kpc, since gas beyond that point is unlikely to be influenced by the central galaxy.  This latter limit affects O~VI column-density predictions for halos of mass $\gtrsim 10^{13} \, M_\odot$ but has negligible effects on smaller systems.

\subsubsection{Radial Profiles}

Figure \ref{fig-5} shows the radial profiles of $N_{\rm OVI} (r_{\rm proj})$ for pNFW models spanning the circular-velocity range $80 \, {\rm km \, s^{-1}} \leq v_{\rm c,max} \leq 350 \, {\rm km \, s^{-1}}$.  The potential wells of all models have an identical shape, with $r_{200} / r_{\rm s} = 10$, and therefore all have $M_{200} = (1.5 \times 10^{12} \, M_\odot) v_{200}^3$, with a mass range $9.6 \times 10^{10} \, M_\odot \leq M_{200} \leq 8.0 \times 10^{12} \, M_\odot$.

Two features stand out:  (1) the column-density profiles are generally flat to beyond 100~kpc, and (2) the characteristic column density is $N_{\rm O VI} \approx 10^{14} \, {\rm cm^{-2}}$ over the entire mass range.  The flatness of the column-density profiles reflects two separate features of the pNFW models.  First, the characteristic electron density profile at small radii is $n_e \propto r^{-1.2}$, as shown in Figure \ref{fig-1}.  Integrating density along lines of sight at a given projected radius therefore tends to give $N_{\rm CGM} \propto r_{\rm proj}^{-0.2}$.  This result is close to the column-density profile slope in the middle column of Figure \ref{fig-5}.  Second, the primary contribution to the total O~VI column density in some cases comes from radii $\gtrsim 100$~kpc, as shown by the black dotted lines in Figure \ref{fig-5}.  This circumstance arises when the temperature-dependent ionization correction for O~VI is more favorable at large radii than at small radii.  In those cases, $N_{\rm O VI}$ is nearly independent of $r_{\rm proj}$ to beyond 100~kpc because it is coming primarily from a thick shell at $\sim 100$~kpc. 

\subsubsection{Scaling with Halo Mass}
\label{sec-NOVI_scaling}

A simple scaling argument captures the essence of the insensitivity of $N_{\rm OVI}$ to halo mass.  The total hydrogen column density along a line of sight through a precipitation-limited CGM is
\begin{eqnarray}
N_{\rm H}   & \, \approx \, & 2 n_e(r_{\rm proj}) \, r_{\rm proj} 
						  \\
		   & \, \approx \, & \frac {2 r_{\rm proj}} {t_{\rm ff}(r_{\rm proj}) }  
						\left[ \frac {3kT} {10  \Lambda(T)}   \right]
					          \label{eq-NH_step2}    
					          \\
			& \, \approx \, & \frac {3} {2^{1/2} 5} 
			                          \left[ \frac {\mu m_p v_c^3} 
							{\Lambda(2 T_\phi)}   \right]
							 \label{eq-NH_step3}    
							\\
			& \, \approx \, & 7 \times 10^{19} \, {\rm cm^{-2}}
					   \left( \frac {Z} {Z_\odot} \right)^{-0.7} v_{200}^{4.7}
					    \label{eq-NH_step4}    
					 \; \; 
\end{eqnarray}
Equation (\ref{eq-NH_step2}) assumes $t_{\rm cool} / t_{\rm ff} = 10$.  Equation (\ref{eq-NH_step3})  sets $T = 2 T_\phi$ in the cooling function, because the CGM temperature at small radii determines the radial structure of the ambient medium.  Equation (\ref{eq-NH_step4})  assumes $\Lambda = 1.2 \times 10^{-22} \, {\rm erg \, cm^3 s^{-1}} (T/10^6 \, {\rm K})^{-0.85}(Z/Z_\odot)^{0.7} $, which approximates the cooling functions of \citet{sd93} in the temperature range $10^{5.5} \, {\rm K} \leq T \leq 10^{6.5} \, {\rm K}$ and the abundance range $0.1 \leq Z/Z_\odot \leq 1.0$.  Converting to an oxygen column density requires an expression for the oxygen abundance.  This calculation assumes O/H = $5.4 \times 10^{-4} (Z/Z_\odot)$ at $v_{\rm c,max} = 200 \, {\rm km \, s^{-1}}$, so that 
\begin{equation}
  N_{\rm O} \, \approx \, 4 \times 10^{16} \, {\rm cm^{-2}}  
  					\left( \frac {Z} {Z_\odot} \right)^{0.3} v_{200}^{4.7}
		\; \; 
\label{eq-N_O}
\end{equation}
The remaining step applies an O~VI ionization correction.  Fitting a power law to the O~VI ionization fractions of \citet{sd93} gives $f_{\rm OVI} = 0.006 (T/10^6 \, {\rm K})^{-2.3}$ in the temperature range $10^{5.5} \, {\rm K} \leq T \leq 10^{6.5} \, {\rm K}$.  Gas at $r \sim 0.5 r_{200}$ and $T \approx T_\phi$ generally contributes the bulk of the O~VI column density, and using a temperature $T = T_\phi$ to determine the O~IV ionization fraction gives
\begin{equation} 
  N_{\rm OVI} \, \approx \, 1 \times 10^{14} \, {\rm cm^{-2}}  
  					\left( \frac {Z} {Z_\odot} \right)^{0.3} v_{200}^{0.1}
  \; \; . 
\end{equation}
This value is indeed close to the characteristic column density of the profiles in Figure \ref{fig-5} and has a negligible dependence on halo mass within the range corresponding to ambient temperatures between $10^{5.5}$~K and $10^{6.5}$~K.  In other words, the halo-mass dependence of total column density in a precipitation-limited CGM ($N_{\rm H} \appropto M_{200}^{1.56}$) almost exactly offsets the steep decline in O~VI ionization fraction ($f_{\rm OVI} \appropto M_{200}^{-1.53}$) within this mass range, while the precipitation condition mitigates the sensitivity of $N_{\rm OVI}$ to metallicity.

At the endpoints of this mass range, the pNFW model predictions assuming pure collisional ionization drop off.  On the high-mass end, the increasing ambient temperature strongly suppresses the O~VI ionization fraction \citep{Oppenheimer_2016MNRAS.460.2157O}.  On the low-mass end, the ambient temperature becomes insufficient to produce observable O~VI lines through collisional ionization.  However, the thermal pressure in the precipitation-limited CGM of a halo with $M_{200} \lesssim 10^{11.5} M_\odot$ is $n_{\rm H} T \lesssim 5 \, {\rm K \, cm^{-3}}$ at $\gtrsim 50$~kpc, which is small enough for the metagalactic ionizing radiation at $z \sim 0$ to boost the O~VI column density above the collisional-ionization prediction \citep[e.g.,][]{Stern_2018arXiv180305446S}.  In that case, equation (\ref{eq-N_O}) gives an upper limit $N_{\rm OVI} \lesssim 10^{14} \, {\rm cm^{-2}} (Z/Z_\odot)^{0.3} (M_{200} / 10^{11} \, M_\odot)^{1.6}$, assuming $f_{\rm OVI} \lesssim 0.2$.

\subsubsection{Ne~VIII and N~V}

\begin{figure}[t]
\begin{center}
\includegraphics[width=3.5in, trim = 0.1in 0.1in 0.0in 0.0in]{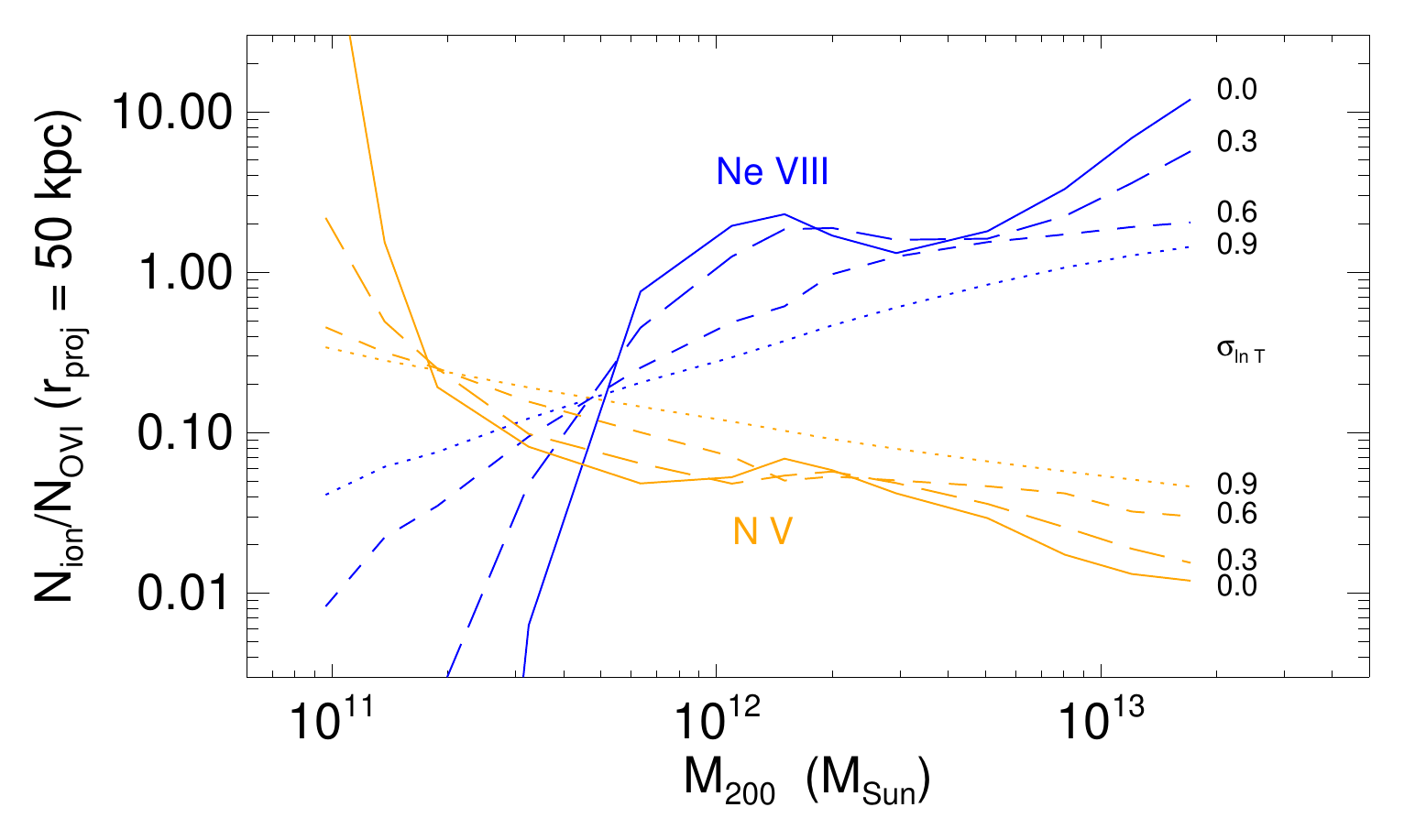} \\
\end{center}
\caption{ \footnotesize 
Column-density ratios predicted by precipitation-limited models for the ambient CGM at a projected radius of 50~kpc .  Blue lines show $N_{\rm NeVIII} / N_{\rm OVI}$ ratios.  Orange lines show $N_{\rm NV} / N_{\rm OVI}$.  The solid lines in each panel show predictions for a static CGM.  The other lines show predictions for a CGM with log-normal temperature fluctuations corresponding to $\sigma_{\ln T} = 0.3$ (long-dashed lines), 0.6 (short-dashed lines), and 0.9 (dotted lines).  
\vspace*{1em}
\label{fig-6}}
\end{figure}

Observations of Ne~VIII and N~V absorption lines can be used to test these models.  The solid lines in Figure \ref{fig-6} show how the ambient $N_{\rm NeVIII}/N_{\rm OVI}$ and $N_{\rm NV}/N_{\rm OVI}$ ratios in a static precipitation-limited CGM depend on halo mass.  At $M_{200} \sim 10^{12} \, M_\odot$, Ne~VIII absorption is predicted to be comparable to O~VI, with $N_{\rm NeVIII} \sim 10^{14} \, {\rm cm^{-2}}$.  However, the predictions for lower halo masses drop sharply because their ambient CGM temperatures are too low for significant Ne~VIII absorption.   In contrast, static pNFW models for $M_{200} \gtrsim 10^{11.5} \, M_\odot$ predict $N_{\rm NV}/N_{\rm OVI} \lesssim 0.1$ and $N_{\rm NeV} \sim 10^{13} \, {\rm cm^{-2}}$.   The other lines in Figure \ref{fig-6} illustrate the dynamic CGM models presented in \S \ref{sec-Dynamic}.

These static-model predictions for $N_{\rm NeVIII}$ and $N_{\rm NV}$ generally agree with the available absorption-line data for the CGM in $10^{12} \, M_\odot$ halos.  Observations of the COS-HALOS galaxies typically fail to detect N~V \citep[e.g.,][]{Werk2016_ApJ...833...54W}, giving mostly upper limits ($N_{\rm NV} \lesssim 10^{13.4-13.8} \, {\rm cm^{-2}}$) and just three detections with $N_{\rm NV}/N_{\rm OVI} \sim 0.1$.  Fewer targets permit observations of Ne~VIII absorption, but the existing detections cluster around $N_{\rm NeVIII} \sim 10^{14} \, {\rm cm^{-2}}$ \citep[e.g.,][]{Pachat_2017MNRAS.471..792P,Frank_2018MNRAS.476.1356F,Burchett_2018arXiv181006560B}.

\subsection{Dynamic CGM}
\label{sec-Dynamic}

Dynamic disturbances in the CGM can alter the absorption-line predictions of precipitation-limited models by perturbing the ionization fractions at each radius.  The typical velocity widths and centroid offsets of O~VI lines from the central galaxy are indeed suggestive of sub-Keplerian disturbances and show that $N_{\rm OVI}$ is positively correlated with line width, as quantified by the Doppler $b$ parameter \citep{Werk2016_ApJ...833...54W}.  Those findings motivate an extension of the pNFW model that allows for temperature fluctuations at each radius in the CGM.

\subsubsection{Temperature Fluctuations}

\begin{figure}[t]
\begin{center}
\includegraphics[width=3.5in, trim = 0.1in 0.1in 0.0in 0.0in]{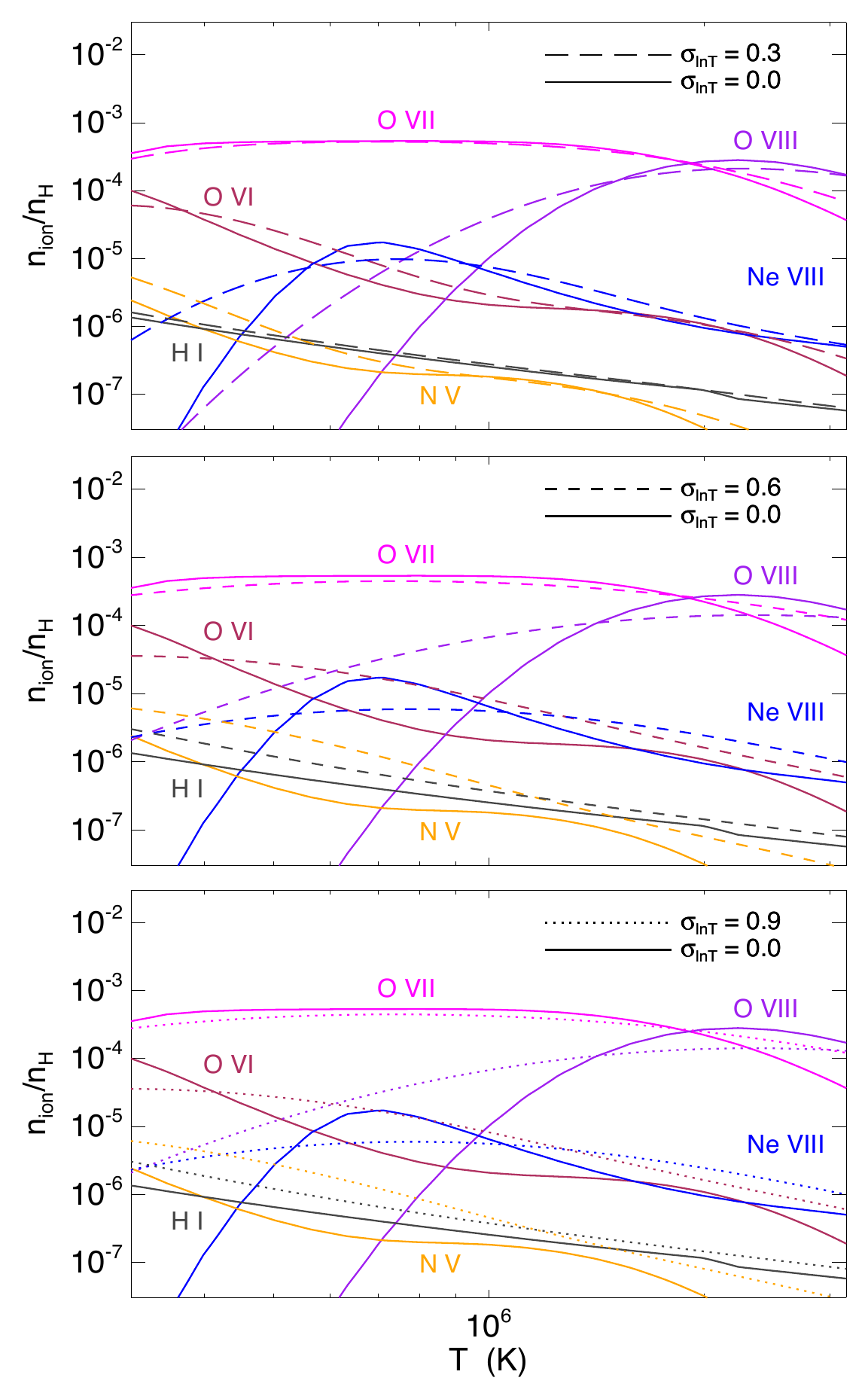} \\
\end{center}
\caption{ \footnotesize 
The effects of temperature fluctuations on mean ion abundances in the CGM.  Solid lines in each panel show ion abundances calculated using ionization fractions from the coronal ionization equilibrium models of \citet{sd93}.  The abundances are assumed to be solar, with $n_{\rm O} / n_{\rm H} = 5 \times 10^{-4}$, $n_{\rm Ne} / n_{\rm H} = 1.3 \times 10^{-4}$, and $n_{\rm N} / n_{\rm H} = 1.1 \times 10^{-4}$.  Long-dashed lines in the top panel show how the mean ion fractions change in the presence of log-normal temperature fluctuations with $\sigma_{\ln T} = 0.3$, assuming local coronal ionization equilibrium.  Short-dashed lines in the middle panel show predictions for $\sigma_{\ln T} = 0.6$.  Dotted lines in the bottom panel show predictions for $\sigma_{\ln T} = 0.9$.
\vspace*{1em}
\label{fig-7}}
\end{figure}

The simplest extension assumes a distribution of gas temperatures having the same log-normal dispersion, $\sigma_{\ln T}$, at all radii \citep[e.g.,][]{Faerman_2017ApJ...835...52F,McQuinnWerk_2018ApJ...852...33M}.  Figure~\ref{fig-7} illustrates how such a dispersion affects the ion fractions when they are convolved with a log-normal temperature distribution, assuming collisional ionization equilibrium remains valid, a critical assumption that will be discussed in \S \ref{sec-CIE}.  If it holds, the distribution of ion fractions at each radius broadens as $\sigma_{\ln T}$ increases, with greater effects on the minority ionization species.  In particular, the O~VI ionization fraction associated with gas at a mean temperature $\approx 10^6 \, {\rm K}$ rises by nearly an order of magnitude as the temperature dispersion approaches $\sigma_{\ln T} \approx 0.9$, causing a substantial increase in $N_{\rm OVI}$ if such a temperature dispersion is present in the CGM around real galaxies.

\begin{figure}[t]
\begin{center}
\includegraphics[width=3.5in, trim = 0.1in 0.1in 0.0in 0.0in]{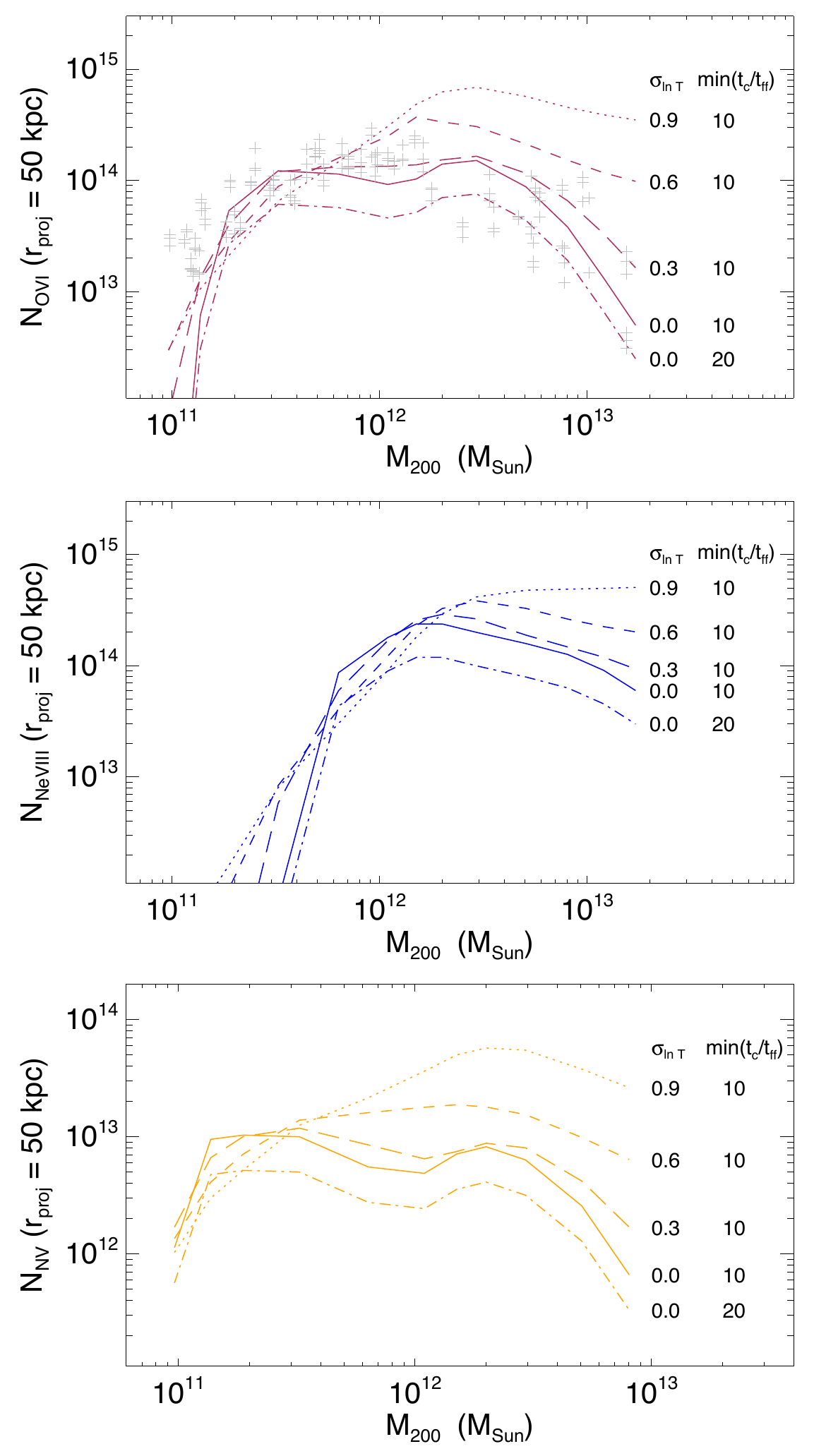} \\
\end{center}
\caption{ \footnotesize 
Column densities at a projected radius of 50~kpc predicted by precipitation-limited models for the ambient CGM of halos in the mass range $10^{11} \, M_\odot$--$10^{13} \, M_\odot$.   Abundances in the CGM are assumed to be solar for $M_{200} = 2 \times 10^{12} \, M_\odot$.   The top panel shows $N_{\rm OVI}$, and crosses in that panel show $N_{\rm OVI}$ measurements of simulated galactic halos from \citet{Oppenheimer_2018MNRAS.474.4740O}.  The middle panel shows $N_{\rm NeVIII}$, and the bottom panel shows $N_{\rm NV}$.
\vspace*{1em}
\label{fig-8}}
\end{figure}

Figure~\ref{fig-8} shows how this extension alters the pNFW model predictions for CGM absorption lines at a projected radius of 50~kpc.  The top panel presents $N_{\rm OVI}$ predictions, along with a set of predictions from the numerical simulations of \citet{Oppenheimer_2018MNRAS.474.4740O}.  Both the pNFW predictions and the simulations feature a broad plateau at $N_{\rm OVI} \approx 10^{14} \, {\rm cm^{-2}}$ in the halo mass range $10^{11} \, M_\odot \lesssim M_{200} \lesssim 10^{13} \, M_\odot$, in accordance with the scaling argument in \S \ref{sec-NOVI_scaling}.  At $M_{200} \lesssim 10^{11} \, M_\odot$, the O~VI predictions rapidly drop, because the ambient CGM temperature is not great enough to produce appreciable quantities of O$^{5+}$.  However, these pNFW models do not account for production of O$^{5+}$ by photoionization, nor do they account for hot galactic outflows that may extend into the CGM at temperatures exceeding the virial temperature.

\begin{figure}[t]
\begin{center}
\includegraphics[width=3.5in, trim = 0.1in 0.1in 0.0in 0.0in]{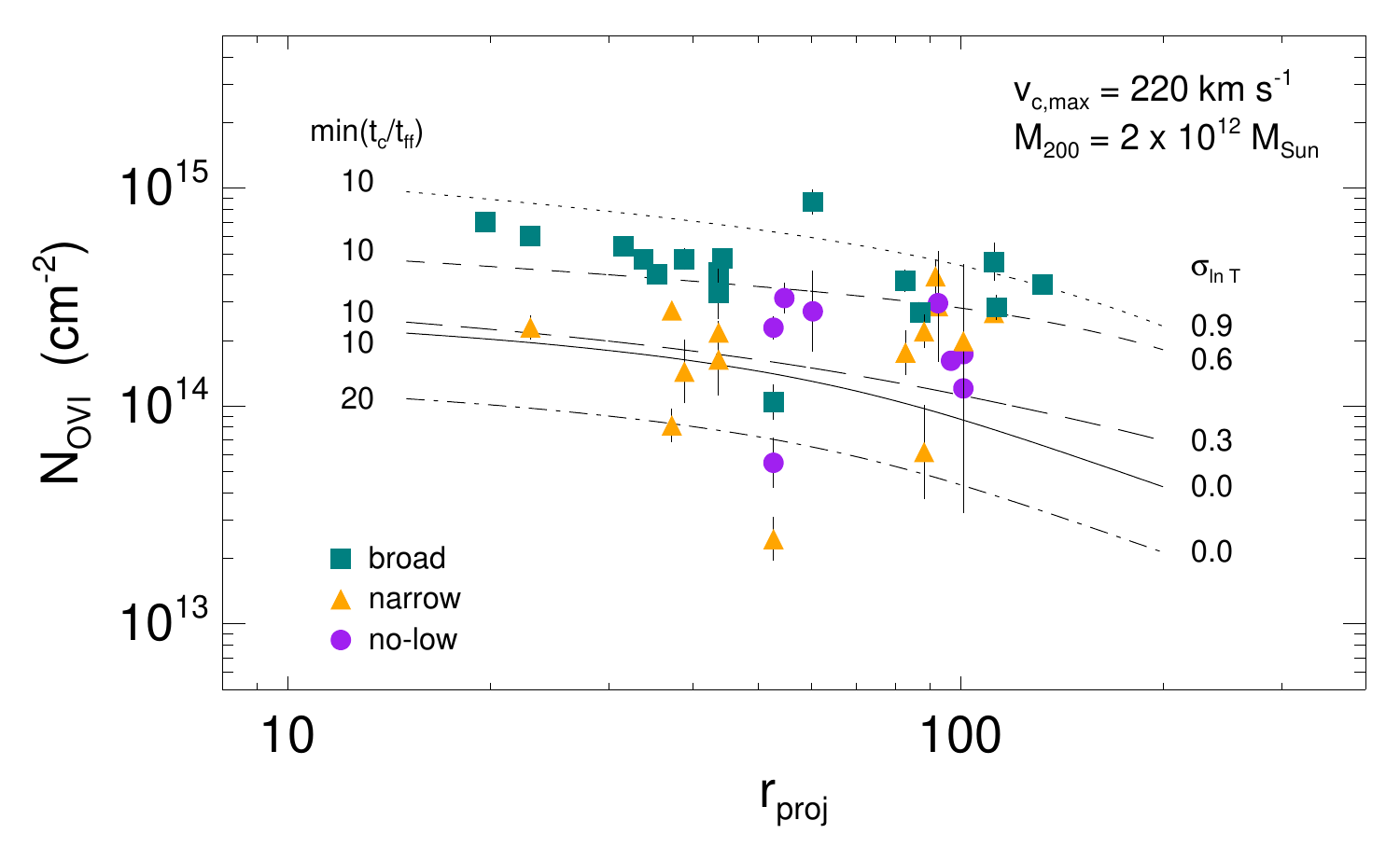} \\
\end{center}
\caption{ \footnotesize 
Comparison of $N_{\rm OVI}$ model predictions with COS-HALOS data from \citet{Werk2016_ApJ...833...54W}.  Lines show column-density profiles for the models shown in Figure~\ref{fig-6} at $v_{\rm c,max} = 220 \, {\rm km \, s^{-1}}$ and $M_{200} = 2 \times 10^{12} \, M_\odot$.  Differences in the data symbols show whether the O~VI absorption line is broad (teal squares) or narrow (orange triangles) or has no associated low-ionization gas (purple circles), as described in \citet{Werk2016_ApJ...833...54W}.  Notice that the broad systems with associated low-ionization absorption are generally consistent with the column-density profiles of precipitation-limited models having $\sigma_{\ln T} \approx 0.7$ and $\min (t_{\rm cool} / t_{\rm ff}) = 10$.
\vspace*{1em}
\label{fig-9}}
\end{figure}

At $M_{200} \gtrsim 10^{11.7} \, M_\odot$, temperature fluctuations substantially enhance the ambient O~VI column density of a precipitation-limited CGM.  Figure~\ref{fig-9} compares those model predictions to a subset of COS-HALOS observations that were analyzed in detail by \citet{Werk2016_ApJ...833...54W}.  They divided those observations into three categories.  Two categories have low-ionization absorption lines coinciding in velocity with the O~VI lines and were divided according to whether the O~VI line was ``broad" ($b > 40 \, {\rm km \, s^{-1}}$) or ``narrow" ($b < 40 \, {\rm km \, s^{-1}}$).  The third category, called ``no-lows," consists solely of O~VI absorption lines without associated low-ionization absorption.  The ``broad" category tends to have the strongest absorbers, with $10^{14.5} \, {\rm cm^{-2}} \lesssim N_{\rm OVI} \lesssim 10^{15} \, {\rm cm^{-2}}$, a level that has been difficult for simulations of the CGM to achieve \citep[e.g.,][]{Hummels_2013MNRAS.430.1548H}.  However, the ``broad" O~VI absorbers agree well with pNFW models having $\sigma_{\ln T} \approx 0.7$, while the ``narrow" absorbers are more consistent with nearly static pNFW models.  

\subsubsection{Adiabatic Uplift}

According to this model, the strongest CGM O~VI absorption lines originate in ambient media with large temperature fluctuations.  Outflows from the central galaxy can produce such fluctuations by lifting low-entropy gas to greater altitudes.  It is not necessary for the uplifted gas to originate within the galactic disk.  As in the cores of galaxy clusters, high-entropy bubbles that buoyantly rise through the ambient medium can lift lower-entropy CGM gas nearly adiabatically, either on their leading edges or within their wakes.

Uplifted gas that remains in pressure balance with its surroundings adiabatically cools as it rises, leading to temperature fluctuations with
\begin{equation}
  \sigma_{\rm \ln T} \approx \frac {3} {5} \sigma_{\ln K}
  \; \; ,
\end{equation}
where $\sigma_{\ln K}$ is the dispersion of entropy fluctuations resulting from uplift.   Persistent temperature fluctuations with $\sigma_{\ln T} \gtrsim 0.6$ therefore imply a distribution of entropy fluctuations with $\sigma_{\ln K} \gtrsim 1$.  In an adiabatic medium with a background profile $K \propto r^{2/3}$, entropy fluctuations of this amplitude can be achieved by lifting CGM gas a factor $\approx e^{3/2} \approx 5$ in radius.

\subsubsection{Internal Gravity Waves}

One way to characterize the effects of CGM uplift is in terms of internal gravity waves, which oscillate at a frequency $\sim t_{\rm ff}^{-1}$.  Internal gravity waves are thermally unstable\footnote{Technically, they are overstable, because they oscillate.} in a thermally balanced medium with an entropy gradient $\alpha_K \equiv d \ln K / d \ln r  \gg ( t_{\rm ff} / t_{\rm cool})^2$.  Their oscillation amplitudes grow on a timescale $\sim t_{\rm cool}$ until they saturate with $\sigma_{\ln K} \sim \alpha_K^{1/2} ( t_{\rm ff} / t_{\rm cool})$ \citep{McCourt+2012MNRAS.419.3319M,ChoudhurySharma_2016MNRAS.457.2554C,Voit_2017_BigPaper}.  Producing precipitation and multiphase gas in such a medium requires a mechanism that drives those oscillations nonlinear and then into overdamping, which leads to condensation.

\citet{Voit_2018arXiv180306036V} recently presented an analysis of circumgalactic precipitation showing that a gravitationally stratified medium with $K \propto r^{2/3}$ and $t_{\rm cool} / t_{\rm ff} \approx 10$ begins to produce condensates when forcing of gravity-wave oscillations causes the velocity dispersion to reach $\sigma_{\rm t} \approx 0.5 \sigma_v$, where $\sigma_v \approx v_c / \sqrt{2}$ is the one-dimensional stellar velocity dispersion corresponding to $v_c$.  When expressed in terms of circular velocity, that critical velocity dispersion is $\sigma_{\rm t} \approx ( 70 \, {\rm km \, s^{-1}} ) v_{200}$, which is equivalent to $b \approx ( 100 \, {\rm km \, s^{-1}} ) v_{200}$ if thermal broadening is negligible.  Gravity waves with that velocity amplitude in a CGM with $t_{\rm cool} / t_{\rm ff} \approx 10$ can no longer be considered adiabatic, because the gas in the low-entropy tail of the resulting entropy distribution has a cooling time comparable to $t_{\rm ff}$.

\subsubsection{Stimulation and Regulation of Condensation}
\label{sec-Condensation}

Another way to view the significance of $\sigma_{\ln T} \gtrsim 0.6$ is in terms of isobaric cooling-time fluctuations, which have  
\begin{equation}
  \sigma_{\ln t_{\rm cool}} \approx (2 - \lambda) \sigma_{\ln T}
\end{equation}
in a medium with $\lambda \equiv d \ln \Lambda / d \ln T$.  In the vicinity of $10^6 \, {\rm K}$, the cooling functions of \citet{sd93} have $\lambda \approx -0.85$, implying $\sigma_{\ln t_{\rm cool}} \gtrsim 1.7$ in a medium with $\sigma_{\ln T} \gtrsim 0.6$.  The low-entropy tail of such a distribution (more than $1 \sigma$ below the mean) has $t_{\rm cool} \lesssim 2 t_{\rm ff}$ if the mean ratio is $t_{\rm cool} / t_{\rm ff} \approx 10$.  The lowest-entropy (shortest cooling-time) gas is therefore susceptible to condensation during a single gravity-wave oscillation.  Larger temperature fluctuations, with $\sigma_{\ln T} \approx 0.9$ and $\sigma_{\ln t_{\rm cool}} \gtrsim 2.6$, imply that gas more than $1 \sigma$ below the mean cooling time has $t_{\rm cool} \lesssim 0.7 t_{\rm ff}$.  In that case, a large fraction of the CGM would cool on a gravitational timescale.

Intriguingly, the ridge line of green squares representing ``broad" O~VI systems in Figure~\ref{fig-9} resides in the region corresponding to pNFW models with $0.6 \lesssim \sigma_{\ln T} \lesssim 0.9$.  According to the preceding argument, this is exactly where forcing of gravity waves in a medium with a mean ratio $t_{\rm cool} / t_{\rm ff} \approx 10$ should drive it into precipitation.  In the framework of precipitation-regulated feedback, the response of the galaxy should be a release of energy that raises the ambient $t_{\rm cool} / t_{\rm ff}$ ratio until it suppresses further precipitation.  Low-ionization condensates might outlive the feedback event, while the CGM settles and the gravity waves damp.  The ``narrow" O~VI systems of \citet{Werk2016_ApJ...833...54W} may be resulting from that damping process.

In the context of those interpretations of ``broad" and ``narrow" O~VI systems, the ``no-lows" would appear to arise from temperature fluctuations associated with gravity waves that are below the threshold for condensation.  As a population, the ``no-lows" have smaller line widths than the ``broad" systems, with a mean $\langle b \rangle \approx 50 \, {\rm km \, s^{-1}}$ and $\max(b) \approx 70 \, {\rm km \, s^{-1}}$.  The ``broad" systems, in contrast, have $\langle b \rangle \approx 90 \, {\rm km \, s^{-1}}$ and $\max(b) \approx 160 \, {\rm km \, s^{-1}}$.  Those characteristics are consistent with the notion that CGM gas within a $\sim 10^{12} \, M_\odot$ halo is driven into condensation when its velocity dispersion approaches $\sigma_{\rm t} \sim 70 \, {\rm km \, s^{-1}}$. 

\subsection{Collisional Ionization Equilibrium}
\label{sec-CIE}

\begin{figure}[t]
\begin{center}
\includegraphics[width=3.5in, trim = 0.1in 0.1in 0.0in 0.0in]{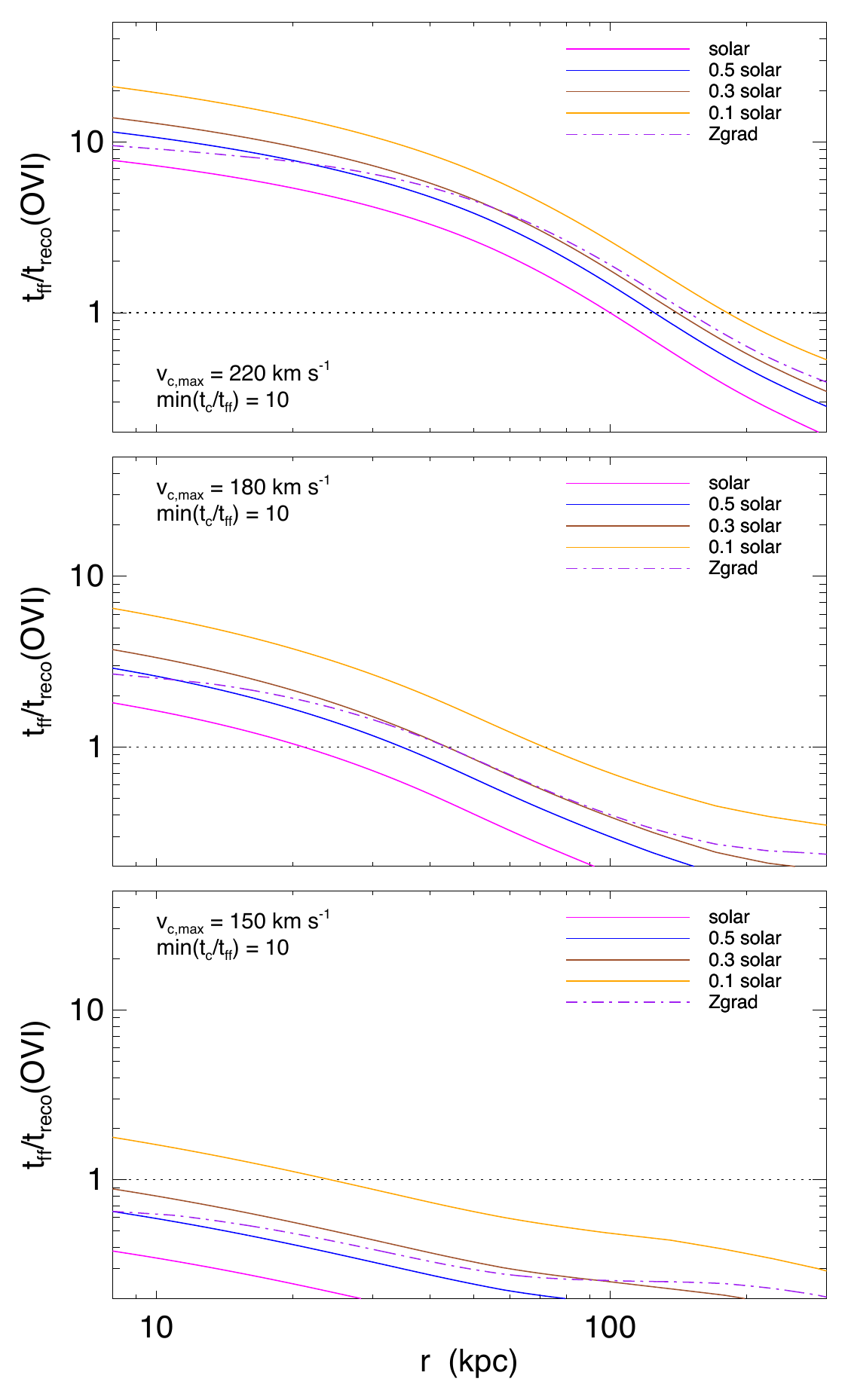} \\
\end{center}
\caption{ \footnotesize 
Ratio of freefall time to O~VI recombination time as a function of radius in pNFW models with $\min (t_{\rm cool} / t_{\rm ff} = 10$ in halos with $v_{\rm c,max} = 220 \, {\rm km \, s^{-1}}$ (top panel), $180 \, {\rm km \, s^{-1}}$ (middle panel), and $150 \, {\rm km \, s^{-1}}$ (bottom panel).  Lines are color-coded as in Figure~\ref{fig-4}.  A horizontal dotted line shows a ratio of unity, above which the O~VI fraction should remain close to its collisional-equilibrium value as gas moves adiabatically at sub-Keplerian speeds.
\vspace*{4em}
\label{fig-10}}
\end{figure}

Interpretations of the strong COS-HALOS O~VI absorbers that rely on temperature fluctuations hinge on the assumption that ionization fractions remain near collisional ionization equilibrium as the CGM temperature fluctuates.  If the fluctuations are produced on a dynamical timescale $\sim t_{\rm ff}$, then this assumption can be checked by comparing $t_{\rm ff}$ with the O~VI recombination time of gas at the CGM's $n_e$ and $T$.  Figure~\ref{fig-10} shows such a comparison as a function of radius for pNFW models with $\min (t_{\rm cool} / t_{\rm ff}) = 10$ and an O~VI recombination coefficient from the fits of Shull \& van Steenberg (1982).  

In a Milky-Way-like halo with $v_{\rm c,max} = 220 \, {\rm km \, s^{-1}}$, the O~VI recombination time is short compared to the dynamical time at $\lesssim 100$~kpc, out to a radius depending on the CGM abundances.  This dependence on abundance arises because a CGM with lower abundances can persist at greater density without violating the precipitation limit.  Within such a halo, the assumption of collisional ionization equilibrium is valid for large-scale motions of CGM gas on a gravitational timescale, including internal gravity waves and slow outflows.  However, it is not valid for temperature fluctuations associated with short-wavelength sound waves or small-scale turbulence.

The bottom two panels show that the assumption of collisional ionization equilibrium becomes more questionable in lower mass halos, because the precipitation-limited gas density at a given radius is substantially smaller.  Consequently, the O~VI recombination time in a halo with $v_c \lesssim 150 \, {\rm km \, s^{-1}}$ is long compared with the dynamical time, implying that the CGM in such a halo might not remain in collisional ionization equilibrium as adiabatic processes change its temperature.  In that case, the O~VI ion fractions would simply reflect the mean temperature of the ambient medium, unless the CGM pressure is low enough for photoionization to determine the O$^{5+}$ fraction.  \citet{Stern_2018arXiv180305446S} have shown that photoionization dominates collisional ionization at $z \sim 0$ in a CGM with thermal pressure $n_{\rm H} T \lesssim 5 \, {\rm K \, cm^{-3}}$.  Precipitation-limited pressures at $\sim 100$~kpc in halos with $M_{200}  \lesssim 10^{11.5} \, M_\odot$ are lower than this threshold (see \S~\ref{sec-NOVI_scaling}), implying that the collisional-ionization assumption is not valid in the outer regions of those lower-mass halos. 

Ambient temperature fluctuations therefore have the most consequential effects on $N_{\rm OVI}$ in systems with $v_{\rm c,max} \gtrsim 180 \, {\rm km \, s^{-1}}$, corresponding to $M_{200} \gtrsim 10^{12} \, M_\odot$.  In that mass range, the response of O~VI ionization to adiabatic cooling on a gravitational timescale is likely to be interesting and relevant.  Coherent uplift of gas with a transverse extent comparable to the radius will then produce large, low-temperature structures in which O$^{5+}$ is enhanced.  If the adiabatic temperature decrease is large enough, then the highest-density regions in those uplifted structures should have cooling times that lead to spatially correlated condensation, as discussed in \S \ref{sec-ColdGas}.

\section{Speculation about Circulation}
\label{sec-SpeculationCirculation}

The observations analyzed in this paper are consistent with models in which energetic feedback heats the CGM, causing the medium to expand without necessarily unbinding it from the galaxy's halo \citep[e.g.,][]{Voit_PrecipReg_2015ApJ...808L..30V}.  According to those models, expansion must drive down the ambient CGM density so that it does not exceed the observed precipitation limit at $\min (t_{\rm cool} / t_{\rm ff}) \approx 10$.  Otherwise, excessive condensation would lead to overproduction of stars.   In such a scenario, the energy supply from the galaxy at the bottom of the potential well drives CGM circulation instead of strong radial outflows that escape the potential well.  This section considers some of the potential implications of O~VI absorption-line phenomenology within that context, showing that the implied supernova energy input can push much of the CGM beyond $r_{200}$, thereby regulating the fraction of baryons that form stars.

\subsection{$N_{\rm OVI}$ and Active Star Formation}

Actively star-forming galaxies are well-known to have O~VI column densities roughly an order of magnitude greater than those around passive galaxies \citep{ChenMulchaey_2009ApJ...701.1219C,Tumlinson_2011Sci...334..948T,Johnson_OVI_2015MNRAS.449.3263J}.  The models presented in this paper, particularly in Figure~\ref{fig-9}, suggest that star formation enhances O~VI absorption because the energetic outflows that star formation propels into the CGM produce temperature fluctuations with $\sigma_{\ln T} \sim 0.7$.  Without a source of energy to cause fluctuations of that magnitude, the ambient CGM within a precipitation-limited halo of mass $10^{12} \, M_\odot \lesssim M_{200} \lesssim 10^{13} \, M_\odot$ should have $N_{\rm OVI} \approx 10^{13.5-14} \, {\rm cm^{-2}}$.  This model prediction is consistent with the detections and upper limits observed around passive galaxies and implies that the greater O~VI columns observed around star-forming galaxies signify circulation.

\subsection{Circulation and Dissipation}

Galactic outflows that lift low-entropy gas without ejecting it from the galaxy's potential well inevitably drive circulation, because the low-entropy gas ultimately sinks back toward the bottom of the potential well.  The rate of energy input required to sustain the level of circulation suggested by the O~VI observations is substantial. For example, consider the CGM of a galaxy like the Milky Way, which has a mass $M_{\rm CGM} \sim 5 \times 10^{10} \, M_\odot$ within $r_{200}$ (see \S \ref{sec-MilkyWay}).  Sustaining CGM circulation with a one-dimensional velocity dispersion $\sigma_{\rm t} \sim 70 \, {\rm km \, s^{-1}}$ and a characteristic circulation length $l_{\rm circ}$ requires a power input
\begin{eqnarray}
  \dot{E}_{\rm circ} & \; \approx \; & 2 \times 10^{41} \, {\rm erg \, s^{-1}} 
  				\left( \frac {M_{\rm CGM}} {5 \times 10^{10} \, M_\odot} \right) 
				\nonumber \\
  		& ~ &	\times
				\left( \frac {\sigma_{\rm t}} {70  \, {\rm km \, s^{-1}}} \right)^3
		                 \left( \frac {l_{\rm circ}} {100  \, {\rm kpc}} \right)^{-1}
		                 \Gamma 
\end{eqnarray}
in order to offset turbulent dissipation of kinetic energy.  In this expression, the quantity $\Gamma$ represents the dimensionless dissipation rate in units of $\sigma_{\rm t} / l_{\rm circ}$ and is of order unity.

This power input is similar in magnitude to the total supernova power of the galaxy ($\approx 3 \times 10^{41} \, {\rm erg \, s^{-1}}$ at a rate of $10^{51} \, {\rm erg}$ per century).  If supernova-driven outflows are indeed responsible for stirring the CGM so that its circulation velocity remains $\sigma_{\rm t} \sim 70 \, {\rm km \, s^{-1}}$, then much of the supernova power generated within the galaxy must dissipate into heat in its CGM.  Clustered supernovae that produce buoyant superbubbles may be required to transport that supernova energy out of the galaxy with the required efficiency \citep[e.g.,][]{Keller_2014MNRAS.442.3013K,Fielding_2018arXiv180708758F}.  Also, the inferred dissipation rate of CGM circulation exceeds the radiative luminosity of the CGM by more than order of magnitude.  For example, integrating over the electron density profiles inferred by \citet{MillerBregman_2013ApJ...770..118M,MillerBregman_2015ApJ...800...14M} gives bolometric luminosity estimates $\lesssim 10^{40} \, {\rm erg \, s^{-1}}$ for the Milky Way's CGM.  

These estimates imply that dissipation of CGM circulation in galaxies like the Milky Way adds heat energy to the CGM faster than it can be radiated away.  The denser, low-entropy fluctuations may still be able to radiate energy fast enough to condense, but higher-entropy regions are likely to be gaining heat as the kinetic energy of CGM circulation dissipates.  If so, then the ambient CGM responds to this entropy input by expanding at approximately constant temperature, and its expansion gently pushes the outer layers of the CGM beyond $r_{200}$.

\subsection{Supernova Feedback and the Precipitation Limit}

Linking the heat input required to gently lift a galaxy's CGM with the galaxy's total output of supernova energy reproduces a scaling relation more commonly associated with galactic winds moving at escape speed.  According to \S \ref{sec-MilkyWay}, a galaxy like the Milky Way must push at least half of the baryons belonging to its halo outside of $r_{200}$ in order to satisfy the precipitation limit.  The amount of energy necessary to lift those ``missing" baryons to such an altitude is $\sim f_{\rm b} M_{200} v_c^2 \sim (2 \times 10^{59} \, {\rm erg}) v_{200}^5$, which is a significant fraction of all the supernova energy that a stellar population with $M_* \approx 7 \times 10^{10} \, M_\odot$ can produce.  More generally, one can define $f_* \equiv M_* / f_{\rm b} M_{200}$ to be a galaxy's stellar baryon fraction and $f_{\rm heat}$ to be the fraction of its supernova energy that is thermalized in the CGM.  Requiring that heat input to lift a majority of the baryonic mass $f_{\rm b} M_{200}$ beyond $r_{200}$ then gives
\begin{eqnarray}
  f_* & \, \approx \, & \frac {v_c^2} {f_{\rm heat} \epsilon_{\rm SN} c^2}  \\
       & \, \approx \, & 0.2 \left( \frac {f_{\rm heat}} {0.5} \right)
				\left( \frac {\epsilon_{\rm SN}} {5 \times 10^{-6}} \right)
				v_{200}^2
				\label{eq-fstar}
				\; \; ,
\end{eqnarray}
where $\epsilon_{\rm SN} \approx 5 \times 10^{-6}$ is the fraction of $M_* c^2$ that ultimately becomes supernova energy.\footnote{The numerical value corresponds to $10^{51} \, {\rm erg}$ of supernova energy per $100 M_\odot$ of star formation.}   Equation (\ref{eq-fstar}) agrees with the Milky Way's stellar mass fraction, given $v_c = 220 \, {\rm km \, s^{-1}}$.  It also yields a dependence of stellar mass on halo mass ($M_* \propto M_{200}^{5/3}$) that aligns with the results of abundance matching in the mass range $10^{11} \, M_\odot \lesssim M_{200} \lesssim 10^{12} \, M_\odot$ \citep[e.g.,][]{Moster_2010ApJ...710..903M}.  

A similar result can be obtained by assuming that all of the accreting baryons ($f_{\rm b} M_{200}$) enter the central galaxy's interstellar medium and fuel star formation that ejects a fraction $\eta/(\eta + 1)$ of the accreted gas, leaving behind a fraction $1/(\eta + 1)$ in the form of stars \citep[see][and references within]{SomervilleDave_2015ARA&A..53...51S}.  If the scaling of the mass-loading factor $\eta$ is determined by requiring SN energy to eject the gas, then $\eta \propto v_c^{-2}$ and $f_* \propto v_c^2$  \citep{Larson_1974MNRAS.169..229L,DekelSilk1986ApJ...303...39D}.  

However, a literal interpretation of the mass-loading scaling argument does not allow for recycling of gas through the CGM.  Instead, it requires galactic winds to unbind a large fraction of a galaxy's baryons from the parent halo, so that they do not return to the central galaxy.  In contrast, the precipitation interpretation simply requires the supernova energy to regulate the recycling rate through subsonic pressure-driven lifting of the CGM.  The precipitation interpretation therefore appears to be in better alignment with observations showing that the speeds of CGM clouds are usually sub-Keplerian \citep[e.g.,][]{Tumlinson_2011Sci...334..948T,Zhu_2014MNRAS.439.3139Z,Huang_2016MNRAS.455.1713H,Borthakur_2016ApJ...833..259B} and simulations showing that a large proportion of the baryons that end up in stars have cycled at least once through the CGM \citep{Oppenheimer_2010MNRAS.406.2325O,Angles-Alcazar_2017MNRAS.470.4698A}.

\subsection{Associated Low-Ionization Gas}
\label{sec-ColdGas}

Many of the intervening O~VI absorption lines in quasar spectra are well-correlated in velocity with H~I lines that have widths indicating a temperature $< 10^5$~K, far below the temperatures at which collisional ionization produces appreciable O$^{5+}$ \citep[e.g.,][]{Tripp_2008ApJS..177...39T,ThomChen_2008ApJS..179...37T}.  If the O~VI absorbing gas is indeed cospatial with such cool H~I gas, then it would have to be photoionized, and therefore at a pressure lower than the pNFW models presented here predict for the CGM in halos of mass $\gtrsim 10^{11.5} \, M_\odot$.  However, most of the O~VI absorbers in the COS-HALOS sample have low-ionization counterparts (e.g., C~II, N~II, Si~II) indicating that the O~VI gas might not be cospatial with the majority of the H~I gas \citep{Werk2016_ApJ...833...54W}.

Circulation that induces CGM precipitation is a potential origin for correlations in both velocity space and physical space among gas components that are not strictly cospatial.  For example, consider an outflow that lifts ambient CGM gas by a factor of a few in radius over a large solid angle.  The column density of uplifted gas would be comparable to the column density of the CGM itself.  In a halo of mass $\sim 10^{12} \, M_\odot$, the adiabatic temperature drop in the uplifted gas would strongly enhance its O$^{5+}$ content, giving $N_{\rm OVI} \gtrsim 10^{14.5} \, {\rm cm^{-2}}$ (\S \ref{sec-Dynamic}).  If the uplift were sufficient to make $t_{\rm cool} \sim t_{\rm ff}$ {\em in the uplifted gas} (see \S \ref{sec-Condensation}), then some of it would condense and enter a state of photoionization equilibrium before the uplifted gas could descend. 

One likely result is ``shattering" of the condensates into fragments of column density $N_{\rm H} \sim 10^{17} \, {\rm cm^{-2}}$.  That is the maximum column density at which the sound crossing time remains less than the radiative cooling time as the gas temperature drops through $\sim 10^5$~K \citep[e.g.,][]{McCourt_2018MNRAS.473.5407M,LiangRemming_2018arXiv180610688L}.  Those fragments would collectively form a ``mist" of low-ionization cloudlets embedded within the O~VI absorber and would co-move with it.   A cloudlet exposed to the metagalactic ionizing radiation at $z \sim 0$ would have a neutral hydrogen fraction $f_{{\rm H}^0} \approx 10^{-5.5} / U$ and column density $N_{\rm HI} \sim 10^{14.5} \, {\rm cm^{-2}} (U/10^{-3})^{-1}$, where the usual ionization parameter $U$ has been scaled to correspond with observations showing $-4 \lesssim \log U \lesssim -2$ in the low-ionization CGM clouds \citep{Stocke_2013ApJ...763..148S,Werk_2014ApJ...792....8W,Keeney_2017ApJS..230....6K}.  

The narrow H~I absorption components associated in velocity with O~VI absorption often have $10^{13.5} \, {\rm cm^{-2}} \lesssim N_{\rm HI} \lesssim 10^{15.5} \, {\rm cm^{-2}}$ \citep[e.g.,][]{Tripp_2008ApJS..177...39T}, and are therefore are consistent with the presence of at least one and perhaps several such low-ionization cloudlets along a line of sight through a larger-scale O~VI absorber.  Many more cloudlets along a given line of sight would produce stronger H~I absorption, but the precipitation model is not yet well-enough developed to predict either the total amount or the longevity of photoionized gas that would result from this condensation process.  Certainly, the total column of low-ionization gas would not be greater than that of the ambient medium from which it originated.  According to equation (\ref{eq-NH_step4}), the upper bound on the column density of low-ionization gas would be $N_{\rm H} \lesssim 10^{20} \, {\rm cm^{-2}}$, independent of projected radius, which accords with the upper bounds on $N_{\rm H}$ inferred from photoionization modeling  \citep{Stocke_2013ApJ...763..148S,Werk_2014ApJ...792....8W,Keeney_2017ApJS..230....6K}.

Photoionized clouds in pressure equilibrium with a hotter ambient medium have ionization levels determined by the ambient pressure.  However, the pressure and density of low-ionization CGM clouds are currently somewhat uncertain because of uncertainties in the metagalactic photoionizing radiation \citep{Shull_UVB_2015ApJ...811....3S,Chen_2017ApJ...842L..19C,Keeney_2017ApJS..230....6K}.  Some recent analyses favor an ionizing background at the high end of the uncertainty range \citep[e.g.][]{Kollmeier_2014ApJ...789L..32K,Viel_2017MNRAS.467L..86V}, resulting in pressures and densities consistent with the ambient pressures predicted by precipitation-limited models.  According to Figure~9 from \citet{Zahedy_2018arXiv180905115Z}, the relationship between gas density and ionization parameter for such a background is $n_{\rm H} \approx 10^{-5.4} \, {\rm cm^{-3}} / U$, giving $n_{\rm H} T \approx 40 \, {\rm K \, cm^{-3}}$ for $\log U \approx -3$ and $T \approx 10^4 \, {\rm K}$.  For comparison, the ambient pressure at 100~kpc in the solar-metallicity pNFW model illustrated in Figure~\ref{fig-1} is $n_{\rm H} T \approx 40  \, {\rm K \, cm^{-3}}$;  it rises to $400  \, {\rm K \, cm^{-3}}$ at $\approx 35$~kpc and drops to $4  \, {\rm K \, cm^{-3}}$ at $\approx 250$~kpc.  Photoionization models of low-ionization CGM clouds with $-4 \lesssim \log U \lesssim -2$ are therefore completely consistent with pressure confinement by a precipitation-limited ambient medium, given current uncertainties in the metagalactic UV background \citep[see also][]{Zahedy_2018arXiv180905115Z}.

\section{Summary}	
\label{sec-Summary}

This paper has derived predictions for absorption-line column densities of O~VI, O~VII, and O~VIII, plus N~V and Ne~VIII, from models in which susceptibility to precipitation limits the ambient density of CGM gas.  Those models were inspired by observations showing that the $t_{\rm cool}/t_{\rm ff}$ ratio in the CGM around very massive galaxies rarely drops much below 10.  Presumably, that lower limit on $t_{\rm cool}/t_{\rm ff}$ arises because ambient gas with a lower ratio is overly prone to condensation and production of cold clouds that accrete onto the galaxy and fuel energetic feedback that raises $t_{\rm cool}$.

Section~\ref{sec-pNFW} presented a prescription for constructing precipitation-limited models of the ambient CGM (i.e. ``pNFW" models) that have declining outer temperature profiles similar to those observed in galaxy clusters and groups.  Those new models are superior to the precipitation-limited models introduced by \citet[][i.e. ``pSIS" models]{Voit2018_LX-T-R}, which predict gas temperatures too hot to be consistent with X-ray observations of both emission and absorption by the Milky Way's CGM.   For the Milky Way, the pNFW models predict a CGM temperature $\gtrsim 2 \times 10^6 \, {\rm K}$ at $\lesssim 40 \, {\rm kpc}$ that declines to $\lesssim 1 \times 10^6 \, {\rm K}$ at $\gtrsim 200 \, {\rm kpc}$, as well as $N_{\rm OVII} \sim N_{\rm OVIII} \sim 10^{16} \, {\rm cm^{-2}}$ for $0.3 \lesssim Z/Z_\odot \lesssim 1.0$.  Both findings are consistent with Milky Way observations.  Given these temperatures and O~VII column densities, the expected O~VI column density of the Milky Way's {\em ambient} CGM is $N_{\rm OVI} \sim 10^{14} \, {\rm cm}^{-2}$.

Section~\ref{sec-MilkyWay} provided further validation of the pNFW models by comparing them with a broad array of multi-wavelength Milky Way data.   Collectively, the data indicate that the Milky Way's CGM has an electron density profile between $n_e \propto r^{-1.2}$ and $n_e \propto r^{-1.5}$ from 10~kpc to 100~kpc, in agreement with the pNFW model predictions.  As shown previously by \citet{MillerBregman_2013ApJ...770..118M,MillerBregman_2015ApJ...800...14M}, combining the X-ray observations with upper limits on the dispersion measure of LMC pulsars places a lower limit of $Z \gtrsim 0.3 Z_\odot$ on the metallicity of the ambient CGM.  The data are most consistent with a CGM having $10 \lesssim \min(t_{\rm cool}/t_{\rm ff}) \lesssim 20$ and a metallicity gradient going from $Z_\odot$ at $\sim 10$~kpc to $0.3 Z_\odot$ at $\sim 200$~kpc, with a total mass $\sim 5 \times 10^{10} \, M_\odot$ inside of $r_{200}$.

Section~\ref{sec-Static} then applied the pNFW model prescription to predict precipitation-limited O~VI column densities for the ambient CGM in halos from $10^{11} \, M_\odot$ to $10^{13} \, M_\odot$, while assuming that the medium is static.  Perhaps surprisingly, those models give $N_{\rm OVI} \approx 10^{14} \, {\rm cm^{-2}}$ across almost the entire mass range, with low sensitivity to metallicity.  The lack of sensitivity to halo mass arises because the rise in total CGM column density with halo mass nearly offsets the decline in the O$^{5+}$ ionization fraction with increasing CGM temperature.  The lack of sensitivity to metallicity arises because the total CGM column density in a precipitation-limited model is greater for lower metallicities.  These static models also predict $N_{\rm NV} \sim 10^{13} \, {\rm cm^{-2}}$ and $N_{\rm NeVIII} \sim 10^{14} \, {\rm cm^{-2}}$ for the CGM in a $\sim 10^{12} \, M_\odot$ halo, in broad agreement with existing observational constraints.

Section~\ref{sec-Dynamic} relaxed the assumption of a static medium and considered the consequences of CGM circulation for O~VI column densities.  Circulation that lifts low-entropy CGM gas to greater altitudes causes adiabatic cooling that can raise the O$^{5+}$ fraction in an ambient medium with a mean temperature $> 10^{5.5}$~K.  Around a galaxy like the Milky Way, circulation that produces isobaric entropy fluctuations with $\sigma_{\ln K} \gtrsim 1$ gives rise to temperature fluctuations with $\sigma_{\ln T} \gtrsim 0.6$ and boosts the O~VI column density to $N_{\rm OVI} \gtrsim 10^{14.5} \, {\rm cm^{-2}}$, as long as the uplifted gas remains close to collisional ionization equilibrium.  The corresponding fluctuations in cooling time have $\sigma_{\ln t_{\rm cool}} \gtrsim 1.7$, implying that the low-entropy tail of the distribution has $t_{\rm cool}/t_{\rm ff} \lesssim 1$, if the mean ratio is $t_{\rm cool}/t_{\rm ff} \approx 10$.  The strongest O~VI absorbers among the COS-HALOS galaxies are therefore plausible examples of CGM systems that circulation has driven into precipitation.  

Section~\ref{sec-SpeculationCirculation} explored what the O~VI absorption-line phenomenology may be telling us, if that interpretation is correct.  Sustaining CGM circulation with $\sigma_{\rm t} \approx 70 \, {\rm km \, s^{-1}}$ on a length scale $\sim 100$~kpc requires a power input comparable to the total supernova power of a galaxy like the Milky Way.  That may be why the CGM around a massive star-forming galaxy ($M_{200} \gtrsim 10^{12} \, M_\odot$) tends to have an O~VI column density exceeding the $10^{13.5-14} \, {\rm cm^{-2}}$ value expected from a static precipitation-limited ambient medium and typically observed around comparably massive galaxies without star formation.  A large cooling flow is not necessarily implied, because much of the O~VI absorption can be coming from gas that uplift has caused to cool adiabatically rather than radiatively.  If radiative cooling then causes a subset of that uplifted gas to condense, it will form small photoionized condensates embedded within a larger collisionally-ionized structure, accounting for the low-ionization absorption lines frequently observed to be associated in velocity with the strongest O~VI lines.

More generally, requiring supernova energy input to expand the ambient CGM in the potential well of a lower-mass galaxy ($M_{200} \lesssim 10^{12} \, M_\odot$), so as to satisfy the precipitation limit, leads to the relation $f_* \approx 0.2 v_{200}^2$.  The same scaling of stellar baryon fraction with circular velocity emerges from feedback models invoking mass-loaded winds driven by supernova energy, but in the precipitation framework those energy-driven outflows do not need to move at escape velocity and unbind gas from the halo.  Instead, they drive dissipative circulation that causes the ambient CGM to expand subsonically, without necessarily becoming unbound.

Several observational tests of the precipitation framework emerge from these models:
\begin{itemize}

\item The most robust prediction is that the cooling time of the ambient CGM at radius $r$ in a precipitation-limited system should rarely, if ever, be smaller than 10 times the freefall time at that radius.  As a consequence, a lower limit on the entropy profile $K(r)$ and an upper limit on the electron-density profile $n_e(r)$ can be calculated from the shape of the potential well within which the CGM resides.  The Appendix provides fitting formulae for those limiting profiles in halos of mass $10^{11} M_\odot \lesssim M_{200} \lesssim 10^{13} M_\odot$ with CGM abundances ranging from $0.1 Z_\odot$ to $Z_\odot$.  Table~\ref{table-Fitting} lists best-fit coefficients corresponding to $\min(t_{\rm cool}/t_{\rm ff}) = 10$, and also $\min(t_{\rm cool}/t_{\rm ff}) = 20$ for a sparser set of halo masses, because $\min(t_{\rm cool}/t_{\rm ff})$ is observed to range from 10 through 20 in higher-mass systems.  (Greater lower limits on $t_{\rm cool}/t_{\rm ff}$ may apply in precipitation-limited systems that are rotating, because rotation at nearly Keplerian speeds significantly reduces the frequency of buoyant oscillations, thereby lengthening the effective dynamical time in the rotating frame.)

\item Ambient temperatures in the central regions of precipitation-limited systems should be $T \approx \mu m_p v_c^2 / 1.2 k \approx (2.4 \times 10^6 \, {\rm K}) v_{200}^2$, because hydrostatic gas at the precipitation limit has $d \ln P / d \ln r \approx d \ln n_e / d \ln r \approx -1.2$.  At larger radii, the gas temperature depends on the outer pressure boundary condition.  Radial profiles of ambient gas temperature and pressure predicted by pNFW models can be calculated from the $K(r)$ and $n_e(r)$ fitting formulae in the Appendix.  X-ray surface brightness predictions for imaging missions currently under development, such as Lynx
and AXIS,
can be derived from the $n_e(r)$ and $T(r)$ profiles for a given CGM metallicity.

\item Out to radii $\sim 100$~kpc, the total hydrogen column density of a precipitation-limited CGM should be nearly independent of projected radius.  Equation (\ref{eq-NH_step4}) predicts $N_{\rm H} \approx 7 \times 10^{19} \, {\rm cm^{-2}} (Z/Z_\odot)^{-0.7} v_{200}^{4.7}$ for a region in which $t_{\rm cool}/t_{\rm ff} \approx 10$ and $kT \approx \mu m_p v_c^2$.  To obtain more precise $N_{\rm H}(r_{\rm proj})$ predictions, one can integrate over the $n_e(r)$ fits in the Appendix at a projected radius $r_{\rm proj}$.

\item Multiplying $N_{\rm H}(r_{\rm proj})$ by the oxygen abundance gives a prediction for the total oxygen column density.  For a region in which $t_{\rm cool}/t_{\rm ff} \approx 10$ and $kT \approx \mu m_p v_c^2$, equation (\ref{eq-N_O}) gives $N_{\rm O} \approx 4 \times 10^{16} \, {\rm cm^{-2}} (Z/Z_\odot)^{0.3} v_{200}^{4.7}$.

\item Assuming collisional ionization equilibrium, one can derive $N_{\rm OVII}(r_{\rm proj})$ and $N_{\rm OVIII}(r_{\rm proj})$ from $N_{\rm O}(r_{\rm proj})$ by applying ionization corrections determined from $T(r)$.  For galaxies like the Milky Way, pNFW models typically predict $N_{\rm OVI} \sim 2 \times 10^{16} \, {\rm cm^{-2}}$ for $\min(t_{\rm cool}/t_{\rm ff}) = 10$ and smaller values for larger $\min(t_{\rm cool}/t_{\rm ff})$.  A spectroscopic X-ray observatory such as ARCUS
would be capable of testing this prediction in the relatively near future \citep{Bregman_ARCUS_2018AAS...23123717B}.

\item The O~VI absorption lines expected from ambient CGM gas in halos of mass $10^{11} M_\odot \lesssim M_{200} \lesssim 10^{13} M_\odot$ are currently observable, because the pNFW models predict $N_{\rm OVI} \gtrsim 10^{13.5} \, {\rm cm^{-2}}$ at nearly all projected radii (see Figure \ref{fig-5}).   The corresponding H~I column density of the ambient medium is an order of magnitude smaller for a CGM metallicity $\sim Z_\odot$ (see Figure \ref{fig-7}). If the medium is essentially static, the widths of those lines will be consistent with thermal broadening at the ambient temperature. 

\item Collisionally ionized gas in the ambient CGM should have $N_{\rm NV} \lesssim 0.1 N_{\rm OVI}$ in halos with $M_{200} \gtrsim 10^{11.5} \, M_\odot$ and $N_{\rm NeVIII} \approx N_{\rm OVI}$ in halos with $10^{11.7} \, M_\odot \lesssim M_{200} \lesssim 10^{12.5} \, {\rm cm^{-2}} \, M_\odot$ (see Figure~\ref{fig-6}).

\item Circulation of CGM gas in halos of mass $\gtrsim 10^{11.7} M_\odot$ should cause $N_{\rm OVI}$ to correlate positively with the line width and/or its offset from the galaxy's systemic velocity, because greater circulation speeds lead to greater fluctuations in specific entropy, temperature, and ionization state (see Figure~\ref{fig-8}).  However, specific predictions for the relationship between line width and $N_{\rm OVI}$ require a more definite model for CGM circulation.

\item Circulation that produces entropy fluctuations large enough for the low-entropy tail of the distribution to have $t_{\rm cool} \lesssim t_{\rm ff}$ will cause condensates to precipitate out of the ambient gas.  \citet{Voit_2018arXiv180306036V} has shown that the threshold for condensation corresponds to a one-dimensional velocity dispersion $\sigma_{\rm t} \approx 0.35 v_c$ in a background medium with $10 \lesssim t_{\rm cool}/t_{\rm ff} \lesssim 20$.  Low-ionization gas resulting from precipitation is therefore expected to have a dispersion of velocity offsets $\sim 70 \, {\rm km \, s^{-1}}$ at $M_{200} \approx 10^{12} \, M_\odot$ and $\sim 120 \, {\rm km \, s^{-1}}$ at $M_{200} \approx 10^{13} \, M_\odot$.

\item The resulting mist of cloudlets will be photoionized by the metagalactic UV background, with an ionization level determined by the ambient CGM pressure, which can be calculated for pNFW models using the fitting formulae in the Appendix.   Around a galaxy like the Milky Way, those models predict $n_{\rm H} T \approx 400 \, {\rm K \, cm^{-2}}$ at 35~kpc, $n_{\rm H} T \approx 40 \, {\rm K \, cm^{-2}}$ at 100~kpc, and $n_{\rm H} T \approx 4 \, {\rm K \, cm^{-2}}$ at 250~kpc, assuming $\min(t_{\rm cool}/t_{\rm ff}) = 10$.  Those pressure predictions drop by a factor of two for $\min(t_{\rm cool}/t_{\rm ff}) = 20$.

\item  In lower-mass halos, the pNFW models predict smaller CGM pressures that may allow photoionization to produce the observed O~VI column densities.  At radii $\sim 100$~kpc in a halo with $M_{200} \lesssim 10^{11.5}$, the predicted CGM pressure is $n_{\rm H} T \lesssim 5 \, {\rm K \, cm^{-2}}$, and O$^{5+}$ is produced mainly by photoionization.  In that limit, the pNFW models predict $N_{\rm OVI} \lesssim 10^{14} \, {\rm cm^{-2}} (Z/Z_\odot)^{0.3} (M_{200}/10^{11} \, M_\odot)^{1.6}$, based on multiplying $N_{\rm O}$ by $f_{\rm OVI} \lesssim 0.2$.

\item The total column density of photoionized condensed gas cannot exceed that of the ambient medium.  Equation (\ref{eq-NH_step4}) therefore places an upper limit of $N_{\rm H} \lesssim 7 \times 10^{19} \, {\rm cm}^{-2} (Z / Z_\odot)^{-0.7} v_{200}^{4.7}$ on the condensed phase, implying a joint dependence on halo mass and metallicity $\appropto Z^{-0.7} M_{200}^{1.6}$.

\end{itemize}

The author would like to thank J. Bregman, G. Bryan, J. Burchett, H.-W. Chen, M. Donahue, M. Gaspari, S. Johnson, N. Murray, B. Nath, B. Oppenheimer, B. O'Shea, M. Peeples, M. Shull, P. Singh, J. Stern, A. Sternberg, J. Stocke, T. Tripp, J. Tumlinson, J. Werk, and F. Zahedy for stimulating and helpful conversations.  Jess Werk and Hsiao-Wen Chen receive extra credit for helpful comments on earlier drafts of the paper.  Partial support for this work was provided by the Chandra Science Center through grant TM8-19006X.

\appendix

\section{Fitting Formulae for pNFW Profiles}

A single power law provides a good fit to pNFW profiles for the CGM in halos with $10^{11} M_\odot \lesssim M_{200} \lesssim  10^{13} M_\odot$:
\begin{equation}
  K(r) = K_1 \left( \frac {r} {1 \, {\rm kpc}} \right)^{\alpha_K}
  \label{eq-fit_K}
  \; \; .
\end{equation}
The electron-density profiles of pNFW profiles in the same mass range correspond more closely to a shallow power law ($n_e \propto r^{-\zeta_1}$ with $\zeta_1 \approx 1.2$) at small radii and a steeper power law ($n_e \propto r^{-\zeta_2}$ with $\zeta_2 \approx 2.3$) at larger radii (see Figure~\ref{fig-1}).  These two limiting power laws can be joined using the fitting formula
\begin{equation}
  n_e(r) = \left\{ 
  			\left[ n_1 \left( \frac {r} {1 \, {\rm kpc}} \right)^{-\zeta_1} \right]^{-2}
  			+ \left[ n_2 \left( \frac {r} {100 \, {\rm kpc}} \right)^{-\zeta_2} \right]^{-2}
			\right\}^{-1/2}
			  \label{eq-fit_ne}
			\; \; .
\end{equation}
Together, fitting formulae (\ref{eq-fit_K}) and (\ref{eq-fit_ne}) determine the temperature profile via $kT(r) = K(r) n_e^{2/3}(r)$ and the thermal-pressure profile via $P = (\mu_e / \mu) K(r) n_e^{5/3} (r)$.  Table~\ref{table-Fitting} gives the best-fitting coefficients for some representative pNFW profiles.


\begin{table}[h]
\centering
\caption{pNFW Fitting Formula Coefficients (for $r_{\rm s} = 0.1 r_{200}$)}
\begin{tabular}{cccccccccc}
\hline
$v_c \, ( {\rm km \, s^{-1}})$ & $M_{200} \, (M_\odot)$ & 
         $\min(t_{\rm cool}/t_{\rm ff})$ & $Z/Z_\odot$ & $K_1$ & $\alpha_K$ & 
	$ n_1 \, ( {\rm cm^{-3}}) $ & $\zeta_1$ & $n_2 \, ( {\rm cm^{-3}}) $ & $\zeta_2$ \\
\hline
 350 &  $8.0 \times 10^{12}$  & 10 &  1.0 & 3.5  &  0.72 
         &  $8.4 \times 10^{-2}$  & 1.2 & $2.9 \times 10^{-4}$ & 2.1 \\
 350 &  $8.0 \times 10^{12}$  & 10 &  0.5 & 2.7  &  0.73 
         &  $1.2 \times 10^{-1}$ & 1.2 & $3.8 \times 10^{-4}$ & 2.1 \\
 350 &  $8.0 \times 10^{12}$  & 10 &  0.3 & 2.4  &  0.74 
         &  $1.5 \times 10^{-1}$ & 1.2 & $4.2 \times 10^{-4}$ & 2.1 \\
 350 &  $8.0 \times 10^{12}$  & 10 &  $Z_{\rm grad}$ & 3.6  &  0.68 
         &  $8.6 \times 10^{-2}$ & 1.1 & $4.1 \times 10^{-4}$ & 2.0 \\
  ~ & ~ & ~ & ~& ~ & ~ & ~ & ~ & ~ \\
 300 &  $5.1 \times 10^{12}$  & 10 &  1.0 & 3.3  &  0.71 
         &  $5.6 \times 10^{-2}$  & 1.2 & $1.7 \times 10^{-4}$ & 2.1 \\
 300 &  $5.1 \times 10^{12}$  & 10 &  0.5 & 2.6  &  0.72 
         &  $8.0 \times 10^{-2}$ & 1.2 & $2.3 \times 10^{-4}$ & 2.1 \\
 300 &  $5.1 \times 10^{12}$  & 10 &  0.3 & 2.3  &  0.73 
         &  $9.6 \times 10^{-2}$ & 1.2 & $2.6 \times 10^{-4}$ & 2.1 \\
 300 &  $5.1 \times 10^{12}$  & 10 &  $Z_{\rm grad}$ & 3.4  &  0.67 
         &  $5.8 \times 10^{-2}$ & 1.1 & $2.6 \times 10^{-4}$ & 2.1 \\
  ~ & ~ & ~ & ~& ~ & ~ & ~ & ~ & ~ \\
 300 &  $5.1 \times 10^{12}$  & 20 &  1.0 & 5.2  &  0.70 
         &  $2.9 \times 10^{-2}$ & 1.1 & $1.0 \times 10^{-4}$ & 2.1 \\
 300 &  $5.1 \times 10^{12}$  & 20 &  0.5 & 4.1  &  0.71 
         &  $4.3 \times 10^{-2}$ & 1.1 & $1.4 \times 10^{-4}$ & 2.1 \\
 300 &  $5.1 \times 10^{12}$  & 20 &  0.3 & 3.6  &  0.71 
         &  $5.1 \times 10^{-2}$ & 1.2 & $1.6 \times 10^{-4}$ & 2.1 \\
 300 &  $5.1 \times 10^{12}$  & 20 &  $Z_{\rm grad}$ & 5.4  &  0.65 
         &  $3.0 \times 10^{-3}$ & 1.1 & $1.6 \times 10^{-4}$ & 2.1 \\
  ~ & ~ & ~ & ~& ~ & ~ & ~ & ~ & ~ \\
 250 &  $2.9 \times 10^{12}$  & 10 &  1.0 & 3.7  &  0.71 
         &  $2.8 \times 10^{-2}$ & 1.2 & $8.1 \times 10^{-5}$ & 2.2 \\
 250 &  $2.9 \times 10^{12}$  & 10 &  0.5 & 2.8  &  0.72 
         &  $4.2 \times 10^{-2}$ & 1.2 & $1.1 \times 10^{-4}$ & 2.2 \\
 250 &  $2.9 \times 10^{12}$  & 10 &  0.3 & 2.4  &  0.72 
         &  $5.1 \times 10^{-3}$ & 1.2 & $1.3 \times 10^{-4}$ & 2.2 \\
 250 &  $2.9 \times 10^{12}$  & 10 &  $Z_{\rm grad}$ & 3.7 &  0.66 
         &  $2.9 \times 10^{-2}$ & 1.1 & $1.3 \times 10^{-4}$ & 2.1 \\
  ~ & ~ & ~ & ~& ~ & ~ & ~ & ~ & ~ \\
 220 &  $2.0 \times 10^{12}$  & 10 &  1.0 & 4.0  &  0.70 
         &  $1.7 \times 10^{-2}$ & 1.2 & $4.2 \times 10^{-5}$ & 2.3 \\
 220 &  $2.0 \times 10^{12}$  & 10 &  0.5 & 3.0  &  0.71 
         &  $2.5 \times 10^{-2}$ & 1.2 & $6.1 \times 10^{-5}$ & 2.2 \\
 220 &  $2.0 \times 10^{12}$  & 10 &  0.3 & 2.6  &  0.71 
         &  $3.1 \times 10^{-2}$ & 1.2 & $7.2 \times 10^{-5}$ & 2.2 \\
 220 &  $2.0 \times 10^{12}$  & 10 &  $Z_{\rm grad}$ & 4.1  &  0.65 
         &  $1.8 \times 10^{-2}$ & 1.1 & $7.4 \times 10^{-5}$ & 2.2 \\
  ~ & ~ & ~ & ~& ~ & ~ & ~ & ~ & ~ \\
 220 &  $2.0 \times 10^{12}$  & 20 &  1.0 & 6.3  &  0.69 
         &  $8.6 \times 10^{-3}$ & 1.1 & $2.3 \times 10^{-5}$ & 2.3 \\
 220 &  $2.0 \times 10^{12}$  & 20 &  0.5 & 4.8  &  0.70 
         &  $1.3 \times 10^{-2}$ & 1.1 & $3.4 \times 10^{-5}$ & 2.2 \\
 220 &  $2.0 \times 10^{12}$  & 20 &  0.3 & 4.1  &  0.70 
         &  $1.6 \times 10^{-2}$ & 1.1 & $4.2 \times 10^{-5}$ & 2.2 \\
 220 &  $2.0 \times 10^{12}$  & 20 &  $Z_{\rm grad}$ & 6.5  &  0.63 
         &  $9.2 \times 10^{-3}$ & 1.1 & $4.3 \times 10^{-4}$ & 2.1 \\
  ~ & ~ & ~ & ~& ~ & ~ & ~ & ~ & ~ \\
 180 &  $1.1 \times 10^{12}$  & 10 &  1.0 & 5.6  &  0.71 
         &  $5.4 \times 10^{-3}$ & 1.2 & $9.6 \times 10^{-6}$ & 2.2 \\
 180 &  $1.1 \times 10^{12}$  & 10 &  0.5 & 4.0  &  0.71 
         &  $8.7 \times 10^{-3}$ & 1.2 & $1.6 \times 10^{-5}$ & 2.2 \\
 180 &  $1.1 \times 10^{12}$  & 10 &  0.3 & 3.4  &  0.71 
         &  $1.1 \times 10^{-2}$ & 1.2 & $2.1 \times 10^{-5}$ & 2.2 \\
 180 &  $1.1 \times 10^{12}$  & 10 &  $Z_{\rm grad}$ & 5.7  &  0.63 
         &  $5.9 \times 10^{-3}$ & 1.1 & $2.3 \times 10^{-5}$ & 2.2 \\
 ~ & ~ & ~ & ~& ~ & ~ & ~ & ~ & ~ \\
 150 &  $6.3 \times 10^{11}$  & 10 &  0.5 & 6.1  &  0.68 
         &  $2.8 \times 10^{-3}$ & 1.2 & $7.4 \times 10^{-6}$ & 2.3 \\
 150 &  $6.3 \times 10^{11}$  & 10 &  0.3 & 4.9  &  0.68 
         &  $3.4 \times 10^{-3}$ & 1.1 & $9.8 \times 10^{-6}$ & 2.2 \\
 150 &  $6.3 \times 10^{11}$  & 10 &  0.1 & 3.0  &  0.69 
         &  $8.1 \times 10^{-3}$ & 1.2 & $1.9 \times 10^{-5}$ & 2.2 \\
          ~ & ~ & ~ & ~& ~ & ~ & ~ & ~ & ~ \\
 120 &  $3.2 \times 10^{11}$  & 10 &  0.5 & 6.1  &  0.69 
         &  $1.4 \times 10^{-3}$ & 1.2 & $2.3 \times 10^{-6}$ & 2.2 \\
 120 &  $3.2 \times 10^{11}$  & 10 &  0.3 & 5.0  &  0.70 
         &  $1.9 \times 10^{-3}$ & 1.2 & $2.9 \times 10^{-6}$ & 2.2 \\
 120 &  $3.2 \times 10^{11}$  & 10 &  0.1 & 3.1  &  0.71 
         &  $3.7 \times 10^{-3}$ & 1.2 & $5.1 \times 10^{-6}$ & 2.2 \\
          ~ & ~ & ~ & ~& ~ & ~ & ~ & ~ & ~ \\
 120 &  $3.2 \times 10^{11}$  & 20 &  0.5 & 9.6  &  0.69 
         &  $7.0 \times 10^{-4}$ & 1.2 & $1.2 \times 10^{-6}$ & 2.2 \\
 120 &  $3.2 \times 10^{11}$  & 20 &  0.3 & 7.9  &  0.70 
         &  $9.5 \times 10^{-4}$ & 1.2 & $1.5 \times 10^{-6}$ & 2.2 \\
 120 &  $3.2 \times 10^{11}$  & 20 &  0.1 & 4.9  &  0.71 
         &  $1.9 \times 10^{-3}$ & 1.2 & $2.7 \times 10^{-6}$ & 2.2 \\
   \hline
\end{tabular}
   \label{table-Fitting}
\end{table}


\bibliographystyle{apj}

\begin{thebibliography}{}
\expandafter\ifx\csname natexlab\endcsname\relax\def\natexlab#1{#1}\fi

\bibitem[{{Anders} \& {Grevesse}(1989)}]{AndersGrevesse_1989GeCoA..53..197A}
{Anders}, E., \& {Grevesse}, N. 1989, \gca, 53, 197

\bibitem[{{Anderson} \& {Bregman}(2010)}]{AndersonBregman_2010ApJ...714..320A}
{Anderson}, M.~E., \& {Bregman}, J.~N. 2010, \apj, 714, 320

\bibitem[{{Anderson} {et~al.}(2015){Anderson}, {Gaspari}, {White}, {Wang}, \&
  {Dai}}]{Anderson_2015MNRAS.449.3806A}
{Anderson}, M.~E., {Gaspari}, M., {White}, S.~D.~M., {Wang}, W., \& {Dai}, X.
  2015, \mnras, 449, 3806

\bibitem[{{Angl{\'e}s-Alc{\'a}zar} {et~al.}(2017){Angl{\'e}s-Alc{\'a}zar},
  {Faucher-Gigu{\`e}re}, {Kere{\v s}}, {Hopkins}, {Quataert}, \&
  {Murray}}]{Angles-Alcazar_2017MNRAS.470.4698A}
{Angl{\'e}s-Alc{\'a}zar}, D., {Faucher-Gigu{\`e}re}, C.-A., {Kere{\v s}}, D.,
  {et~al.} 2017, \mnras, 470, 4698

\bibitem[{{Asplund} {et~al.}(2004){Asplund}, {Grevesse}, {Sauval}, {Allende
  Prieto}, \& {Kiselman}}]{Asplund_2004A&A...417..751A}
{Asplund}, M., {Grevesse}, N., {Sauval}, A.~J., {Allende Prieto}, C., \&
  {Kiselman}, D. 2004, \aap, 417, 751

\bibitem[{{Basu} \& {Antia}(2008)}]{BasuAntia_2008PhR...457..217B}
{Basu}, S., \& {Antia}, H.~M. 2008, \physrep, 457, 217

\bibitem[{{Borthakur} {et~al.}(2016){Borthakur}, {Heckman}, {Tumlinson},
  {Bordoloi}, {Kauffmann}, {Catinella}, {Schiminovich}, {Dav{\'e}}, {Moran}, \&
  {Saintonge}}]{Borthakur_2016ApJ...833..259B}
{Borthakur}, S., {Heckman}, T., {Tumlinson}, J., {et~al.} 2016, \apj, 833, 259

\bibitem[{{Bregman} {et~al.}(2018){Bregman}, {Hodges-Kluck}, {Li}, {Li},
  {Miller}, \& {Qu}}]{Bregman_ARCUS_2018AAS...23123717B}
{Bregman}, J.~N., {Hodges-Kluck}, E., {Li}, J., {et~al.} 2018, in American
  Astronomical Society Meeting Abstracts, Vol. 231, American Astronomical
  Society Meeting Abstracts \#231, 237.17

\bibitem[{{Bregman} \&
  {Lloyd-Davies}(2007)}]{BregmanLloydDavies_2007ApJ...669..990B}
{Bregman}, J.~N., \& {Lloyd-Davies}, E.~J. 2007, \apj, 669, 990

\bibitem[{{Burchett} {et~al.}(2018){Burchett}, {Tripp}, {Prochaska}, {Werk},
  {Tumlinson}, {Howk}, {Willmer}, {Lehner}, {Meiring}, {Bowen}, {Bordoloi},
  {Peeples}, {Jenkins}, {O'Meara}, {Tejos}, \&
  {Katz}}]{Burchett_2018arXiv181006560B}
{Burchett}, J.~N., {Tripp}, T.~M., {Prochaska}, J.~X., {et~al.} 2018, ArXiv
  e-prints, arXiv:1810.06560

\bibitem[{{Caffau} {et~al.}(2015){Caffau}, {Ludwig}, {Steffen}, {Livingston},
  {Bonifacio}, {Malherbe}, {Doerr}, \& {Schmidt}}]{Caffau_2015A&A...579A..88C}
{Caffau}, E., {Ludwig}, H.-G., {Steffen}, M., {et~al.} 2015, \aap, 579, A88

\bibitem[{{Chen} {et~al.}(2017){Chen}, {Johnson}, {Zahedy}, {Rauch}, \&
  {Mulchaey}}]{Chen_2017ApJ...842L..19C}
{Chen}, H.-W., {Johnson}, S.~D., {Zahedy}, F.~S., {Rauch}, M., \& {Mulchaey},
  J.~S. 2017, \apjl, 842, L19

\bibitem[{{Chen} \& {Mulchaey}(2009)}]{ChenMulchaey_2009ApJ...701.1219C}
{Chen}, H.-W., \& {Mulchaey}, J.~S. 2009, \apj, 701, 1219

\bibitem[{{Choudhury} \& {Sharma}(2016)}]{ChoudhurySharma_2016MNRAS.457.2554C}
{Choudhury}, P.~P., \& {Sharma}, P. 2016, \mnras, 457, 2554

\bibitem[{{Cowie} {et~al.}(1980){Cowie}, {Fabian}, \&
  {Nulsen}}]{Cowie_1980MNRAS.191..399C}
{Cowie}, L.~L., {Fabian}, A.~C., \& {Nulsen}, P.~E.~J. 1980, \mnras, 191, 399

\bibitem[{{Dekel} \& {Silk}(1986)}]{DekelSilk1986ApJ...303...39D}
{Dekel}, A., \& {Silk}, J. 1986, \apj, 303, 39

\bibitem[{{Faerman} {et~al.}(2017){Faerman}, {Sternberg}, \&
  {McKee}}]{Faerman_2017ApJ...835...52F}
{Faerman}, Y., {Sternberg}, A., \& {McKee}, C.~F. 2017, \apj, 835, 52

\bibitem[{{Fang} {et~al.}(2015){Fang}, {Buote}, {Bullock}, \&
  {Ma}}]{Fang_2015ApJS..217...21F}
{Fang}, T., {Buote}, D., {Bullock}, J., \& {Ma}, R. 2015, \apjs, 217, 21

\bibitem[{{Fang} {et~al.}(2006){Fang}, {Mckee}, {Canizares}, \&
  {Wolfire}}]{Fang_2006ApJ...644..174F}
{Fang}, T., {Mckee}, C.~F., {Canizares}, C.~R., \& {Wolfire}, M. 2006, \apj,
  644, 174

\bibitem[{{Fielding} {et~al.}(2018){Fielding}, {Quataert}, \&
  {Martizzi}}]{Fielding_2018arXiv180708758F}
{Fielding}, D., {Quataert}, E., \& {Martizzi}, D. 2018, ArXiv e-prints,
  arXiv:1807.08758

\bibitem[{{Fielding} {et~al.}(2017){Fielding}, {Quataert}, {McCourt}, \&
  {Thompson}}]{Fielding_2017MNRAS.466.3810F}
{Fielding}, D., {Quataert}, E., {McCourt}, M., \& {Thompson}, T.~A. 2017,
  \mnras, 466, 3810

\bibitem[{{Frank} {et~al.}(2018){Frank}, {Pieri}, {Mathur}, {Danforth}, \&
  {Shull}}]{Frank_2018MNRAS.476.1356F}
{Frank}, S., {Pieri}, M.~M., {Mathur}, S., {Danforth}, C.~W., \& {Shull}, J.~M.
  2018, \mnras, 476, 1356

\bibitem[{{Gaspari} {et~al.}(2013){Gaspari}, {Ruszkowski}, \&
  {Oh}}]{Gaspari+2013MNRAS.432.3401G}
{Gaspari}, M., {Ruszkowski}, M., \& {Oh}, S.~P. 2013, \mnras, 432, 3401

\bibitem[{{Gaspari} {et~al.}(2012){Gaspari}, {Ruszkowski}, \&
  {Sharma}}]{Gaspari+2012ApJ...746...94G}
{Gaspari}, M., {Ruszkowski}, M., \& {Sharma}, P. 2012, \apj, 746, 94

\bibitem[{{Gaspari} {et~al.}(2017){Gaspari}, {Temi}, \&
  {Brighenti}}]{Gaspari_2017MNRAS.466..677G}
{Gaspari}, M., {Temi}, P., \& {Brighenti}, F. 2017, \mnras, 466, 677

\bibitem[{{Gatto} {et~al.}(2013){Gatto}, {Fraternali}, {Read}, {Marinacci},
  {Lux}, \& {Walch}}]{Gatto_2013MNRAS.433.2749G}
{Gatto}, A., {Fraternali}, F., {Read}, J.~I., {et~al.} 2013, \mnras, 433, 2749

\bibitem[{{Ghirardini} {et~al.}(2018){Ghirardini}, {Eckert}, {Ettori},
  {Pointecouteau}, {Molendi}, {Gaspari}, {Rossetti}, {De Grandi}, {Roncarelli},
  {Bourdin}, {Mazzotta}, {Rasia}, \& {Vazza}}]{Ghirardini_2018arXiv180500042G}
{Ghirardini}, V., {Eckert}, D., {Ettori}, S., {et~al.} 2018, ArXiv e-prints,
  arXiv:1805.00042

\bibitem[{{Grcevich} \& {Putman}(2009)}]{Grcevich_2009ApJ...696..385G}
{Grcevich}, J., \& {Putman}, M.~E. 2009, \apj, 696, 385

\bibitem[{{Gupta} {et~al.}(2009){Gupta}, {Galeazzi}, {Koutroumpa}, {Smith}, \&
  {Lallement}}]{Gupta_2009ApJ...707..644G}
{Gupta}, A., {Galeazzi}, M., {Koutroumpa}, D., {Smith}, R., \& {Lallement}, R.
  2009, \apj, 707, 644

\bibitem[{{Gupta} {et~al.}(2012){Gupta}, {Mathur}, {Krongold}, {Nicastro}, \&
  {Galeazzi}}]{Gupta_2012ApJ...756L...8G}
{Gupta}, A., {Mathur}, S., {Krongold}, Y., {Nicastro}, F., \& {Galeazzi}, M.
  2012, \apjl, 756, L8

\bibitem[{{Henley} \& {Shelton}(2012)}]{HenleyShelton_2012ApJS..202...14H}
{Henley}, D.~B., \& {Shelton}, R.~L. 2012, \apjs, 202, 14

\bibitem[{{Henley} \& {Shelton}(2013)}]{HenleyShelton_2013ApJ...773...92H}
---. 2013, \apj, 773, 92

\bibitem[{{Hogan} {et~al.}(2017){Hogan}, {McNamara}, {Pulido}, {Nulsen},
  {Vantyghem}, {Russell}, {Edge}, {Babyk}, {Main}, \&
  {McDonald}}]{Hogan_2017_tctff}
{Hogan}, M.~T., {McNamara}, B.~R., {Pulido}, F., {et~al.} 2017, ArXiv e-prints,
  arXiv:1704.00011

\bibitem[{{Huang} {et~al.}(2016){Huang}, {Chen}, {Johnson}, \&
  {Weiner}}]{Huang_2016MNRAS.455.1713H}
{Huang}, Y.-H., {Chen}, H.-W., {Johnson}, S.~D., \& {Weiner}, B.~J. 2016,
  \mnras, 455, 1713

\bibitem[{{Hummels} {et~al.}(2013){Hummels}, {Bryan}, {Smith}, \&
  {Turk}}]{Hummels_2013MNRAS.430.1548H}
{Hummels}, C.~B., {Bryan}, G.~L., {Smith}, B.~D., \& {Turk}, M.~J. 2013,
  \mnras, 430, 1548

\bibitem[{{Jenkins} \& {Shaya}(1979)}]{JenkinsShaya_1979ApJ...231...55J}
{Jenkins}, E.~B., \& {Shaya}, E.~J. 1979, \apj, 231, 55

\bibitem[{{Jenkins} \& {Tripp}(2011)}]{JenkinsTripp_2011ApJ...734...65J}
{Jenkins}, E.~B., \& {Tripp}, T.~M. 2011, \apj, 734, 65

\bibitem[{{Ji} {et~al.}(2018){Ji}, {Oh}, \& {McCourt}}]{Ji_2018MNRAS.476..852J}
{Ji}, S., {Oh}, S.~P., \& {McCourt}, M. 2018, \mnras, 476, 852

\bibitem[{{Johnson} {et~al.}(2015){Johnson}, {Chen}, \&
  {Mulchaey}}]{Johnson_OVI_2015MNRAS.449.3263J}
{Johnson}, S.~D., {Chen}, H.-W., \& {Mulchaey}, J.~S. 2015, \mnras, 449, 3263

\bibitem[{{Keeney} {et~al.}(2017){Keeney}, {Stocke}, {Danforth}, {Shull},
  {Pratt}, {Froning}, {Green}, {Penton}, \&
  {Savage}}]{Keeney_2017ApJS..230....6K}
{Keeney}, B.~A., {Stocke}, J.~T., {Danforth}, C.~W., {et~al.} 2017, \apjs, 230,
  6

\bibitem[{{Keller} {et~al.}(2014){Keller}, {Wadsley}, {Benincasa}, \&
  {Couchman}}]{Keller_2014MNRAS.442.3013K}
{Keller}, B.~W., {Wadsley}, J., {Benincasa}, S.~M., \& {Couchman}, H.~M.~P.
  2014, \mnras, 442, 3013

\bibitem[{{Kollmeier} {et~al.}(2014){Kollmeier}, {Weinberg}, {Oppenheimer},
  {Haardt}, {Katz}, {Dav{\'e}}, {Fardal}, {Madau}, {Danforth}, {Ford},
  {Peeples}, \& {McEwen}}]{Kollmeier_2014ApJ...789L..32K}
{Kollmeier}, J.~A., {Weinberg}, D.~H., {Oppenheimer}, B.~D., {et~al.} 2014,
  \apjl, 789, L32

\bibitem[{{Kuntz} \& {Snowden}(2000)}]{KuntzSnowden_2000ApJ...543..195K}
{Kuntz}, K.~D., \& {Snowden}, S.~L. 2000, \apj, 543, 195

\bibitem[{{Larson}(1974)}]{Larson_1974MNRAS.169..229L}
{Larson}, R.~B. 1974, \mnras, 169, 229

\bibitem[{{Lau} {et~al.}(2009){Lau}, {Kravtsov}, \& {Nagai}}]{lkn09}
{Lau}, E.~T., {Kravtsov}, A.~V., \& {Nagai}, D. 2009, \apj, 705, 1129

\bibitem[{{Li} {et~al.}(2015){Li}, {Bryan}, {Ruszkowski}, {Voit}, {O'Shea}, \&
  {Donahue}}]{Li_2015ApJ...811...73L}
{Li}, Y., {Bryan}, G.~L., {Ruszkowski}, M., {et~al.} 2015, \apj, 811, 73

\bibitem[{{Liang} \& {Remming}(2018)}]{LiangRemming_2018arXiv180610688L}
{Liang}, C.~J., \& {Remming}, I.~S. 2018, ArXiv e-prints, arXiv:1806.10688

\bibitem[{{Maller} \& {Bullock}(2004)}]{MallerBullock_2004MNRAS.355..694M}
{Maller}, A.~H., \& {Bullock}, J.~S. 2004, \mnras, 355, 694

\bibitem[{{McCammon} {et~al.}(2002){McCammon}, {Almy}, {Apodaca}, {Bergmann
  Tiest}, {Cui}, {Deiker}, {Galeazzi}, {Juda}, {Lesser}, {Mihara},
  {Morgenthaler}, {Sanders}, {Zhang}, {Figueroa-Feliciano}, {Kelley},
  {Moseley}, {Mushotzky}, {Porter}, {Stahle}, \&
  {Szymkowiak}}]{McCammon_2002ApJ...576..188M}
{McCammon}, D., {Almy}, R., {Apodaca}, E., {et~al.} 2002, \apj, 576, 188

\bibitem[{{McCourt} {et~al.}(2018){McCourt}, {Oh}, {O'Leary}, \&
  {Madigan}}]{McCourt_2018MNRAS.473.5407M}
{McCourt}, M., {Oh}, S.~P., {O'Leary}, R., \& {Madigan}, A.-M. 2018, \mnras,
  473, 5407

\bibitem[{{McCourt} {et~al.}(2012){McCourt}, {Sharma}, {Quataert}, \&
  {Parrish}}]{McCourt+2012MNRAS.419.3319M}
{McCourt}, M., {Sharma}, P., {Quataert}, E., \& {Parrish}, I.~J. 2012, \mnras,
  419, 3319

\bibitem[{{McNamara} \& {Nulsen}(2007)}]{mn07}
{McNamara}, B.~R., \& {Nulsen}, P.~E.~J. 2007, \araa, 45, 117

\bibitem[{{McNamara} \& {Nulsen}(2012)}]{McNamaraNulsen2012NJPh...14e5023M}
---. 2012, New Journal of Physics, 14, 055023

\bibitem[{{McNamara} {et~al.}(2016){McNamara}, {Russell}, {Nulsen}, {Hogan},
  {Fabian}, {Pulido}, \& {Edge}}]{McNamara_2016arXiv160404629M}
{McNamara}, B.~R., {Russell}, H.~R., {Nulsen}, P.~E.~J., {et~al.} 2016, ArXiv
  e-prints, arXiv:1604.04629

\bibitem[{{McQuinn} \& {Werk}(2018)}]{McQuinnWerk_2018ApJ...852...33M}
{McQuinn}, M., \& {Werk}, J.~K. 2018, \apj, 852, 33

\bibitem[{{Miller} \& {Bregman}(2013)}]{MillerBregman_2013ApJ...770..118M}
{Miller}, M.~J., \& {Bregman}, J.~N. 2013, \apj, 770, 118

\bibitem[{{Miller} \& {Bregman}(2015)}]{MillerBregman_2015ApJ...800...14M}
---. 2015, \apj, 800, 14

\bibitem[{{Moster} {et~al.}(2010){Moster}, {Somerville}, {Maulbetsch}, {van den
  Bosch}, {Macci{\`o}}, {Naab}, \& {Oser}}]{Moster_2010ApJ...710..903M}
{Moster}, B.~P., {Somerville}, R.~S., {Maulbetsch}, C., {et~al.} 2010, \apj,
  710, 903

\bibitem[{{Navarro} {et~al.}(1997){Navarro}, {Frenk}, \& {White}}]{nfw97}
{Navarro}, J.~F., {Frenk}, C.~S., \& {White}, S.~D.~M. 1997, \apj, 490, 493

\bibitem[{{Nulsen}(1986)}]{Nulsen_1986MNRAS.221..377N}
{Nulsen}, P.~E.~J. 1986, \mnras, 221, 377

\bibitem[{{Oppenheimer} {et~al.}(2010){Oppenheimer}, {Dav{\'e}}, {Kere{\v s}},
  {Fardal}, {Katz}, {Kollmeier}, \&
  {Weinberg}}]{Oppenheimer_2010MNRAS.406.2325O}
{Oppenheimer}, B.~D., {Dav{\'e}}, R., {Kere{\v s}}, D., {et~al.} 2010, \mnras,
  406, 2325

\bibitem[{{Oppenheimer} {et~al.}(2018){Oppenheimer}, {Segers}, {Schaye},
  {Richings}, \& {Crain}}]{Oppenheimer_2018MNRAS.474.4740O}
{Oppenheimer}, B.~D., {Segers}, M., {Schaye}, J., {Richings}, A.~J., \&
  {Crain}, R.~A. 2018, \mnras, 474, 4740

\bibitem[{{Oppenheimer} {et~al.}(2016){Oppenheimer}, {Crain}, {Schaye},
  {Rahmati}, {Richings}, {Trayford}, {Tumlinson}, {Bower}, {Schaller}, \&
  {Theuns}}]{Oppenheimer_2016MNRAS.460.2157O}
{Oppenheimer}, B.~D., {Crain}, R.~A., {Schaye}, J., {et~al.} 2016, \mnras, 460,
  2157

\bibitem[{{Pachat} {et~al.}(2017){Pachat}, {Narayanan}, {Khaire}, {Savage},
  {Muzahid}, \& {Wakker}}]{Pachat_2017MNRAS.471..792P}
{Pachat}, S., {Narayanan}, A., {Khaire}, V., {et~al.} 2017, \mnras, 471, 792

\bibitem[{{Pizzolato} \& {Soker}(2005)}]{ps05}
{Pizzolato}, F., \& {Soker}, N. 2005, \apj, 632, 821

\bibitem[{{Planck Collaboration} {et~al.}(2013){Planck Collaboration}, {Ade},
  {Aghanim}, {Arnaud}, {Ashdown}, {Atrio-Barandela}, {Aumont}, {Baccigalupi},
  {Balbi}, {Banday}, \& et~al.}]{Planck_LRGstacks_2013A&A...557A..52P}
{Planck Collaboration}, {Ade}, P.~A.~R., {Aghanim}, N., {et~al.} 2013, \aap,
  557, A52

\bibitem[{{Prasad} {et~al.}(2015){Prasad}, {Sharma}, \&
  {Babul}}]{Prasad_2015ApJ...811..108P}
{Prasad}, D., {Sharma}, P., \& {Babul}, A. 2015, \apj, 811, 108

\bibitem[{{Putman} {et~al.}(2012){Putman}, {Peek}, \&
  {Joung}}]{Putman_2012ARA&A..50..491P}
{Putman}, M.~E., {Peek}, J.~E.~G., \& {Joung}, M.~R. 2012, \araa, 50, 491

\bibitem[{{Salem} {et~al.}(2015){Salem}, {Besla}, {Bryan}, {Putman}, {van der
  Marel}, \& {Tonnesen}}]{Salem_2015ApJ...815...77S}
{Salem}, M., {Besla}, G., {Bryan}, G., {et~al.} 2015, \apj, 815, 77

\bibitem[{{Savage} {et~al.}(2011{\natexlab{a}}){Savage}, {Lehner}, \&
  {Narayanan}}]{Savage2011_OVI_NeVIII_ApJ...743..180S}
{Savage}, B.~D., {Lehner}, N., \& {Narayanan}, A. 2011{\natexlab{a}}, \apj,
  743, 180

\bibitem[{{Savage} {et~al.}(2011{\natexlab{b}}){Savage}, {Narayanan}, {Lehner},
  \& {Wakker}}]{Savage_BroadOVI_2011ApJ...731...14S}
{Savage}, B.~D., {Narayanan}, A., {Lehner}, N., \& {Wakker}, B.~P.
  2011{\natexlab{b}}, \apj, 731, 14

\bibitem[{{Sharma} {et~al.}(2012){Sharma}, {McCourt}, {Quataert}, \&
  {Parrish}}]{Sharma_2012MNRAS.420.3174S}
{Sharma}, P., {McCourt}, M., {Quataert}, E., \& {Parrish}, I.~J. 2012, \mnras,
  420, 3174

\bibitem[{{Shull} {et~al.}(2015){Shull}, {Moloney}, {Danforth}, \&
  {Tilton}}]{Shull_UVB_2015ApJ...811....3S}
{Shull}, J.~M., {Moloney}, J., {Danforth}, C.~W., \& {Tilton}, E.~M. 2015,
  \apj, 811, 3

\bibitem[{{Singh} {et~al.}(2018){Singh}, {Majumdar}, {Nath}, \&
  {Silk}}]{Singh_stacks_2018MNRAS.478.2909S}
{Singh}, P., {Majumdar}, S., {Nath}, B.~B., \& {Silk}, J. 2018, \mnras, 478,
  2909

\bibitem[{{Somerville} \&
  {Dav{\'e}}(2015)}]{SomervilleDave_2015ARA&A..53...51S}
{Somerville}, R.~S., \& {Dav{\'e}}, R. 2015, \araa, 53, 51

\bibitem[{{Stanimirovi{\'c}} {et~al.}(2002){Stanimirovi{\'c}}, {Dickey}, {Kr{\v
  c}o}, \& {Brooks}}]{Stanimirovich_2002ApJ...576..773S}
{Stanimirovi{\'c}}, S., {Dickey}, J.~M., {Kr{\v c}o}, M., \& {Brooks}, A.~M.
  2002, \apj, 576, 773

\bibitem[{{Stern} {et~al.}(2018){Stern}, {Faucher-Gigu{\`e}re}, {Hennawi},
  {Hafen}, {Johnson}, \& {Fielding}}]{Stern_2018arXiv180305446S}
{Stern}, J., {Faucher-Gigu{\`e}re}, C.-A., {Hennawi}, J.~F., {et~al.} 2018,
  ArXiv e-prints, arXiv:1803.05446

\bibitem[{{Stocke} {et~al.}(2013){Stocke}, {Keeney}, {Danforth}, {Shull},
  {Froning}, {Green}, {Penton}, \& {Savage}}]{Stocke_2013ApJ...763..148S}
{Stocke}, J.~T., {Keeney}, B.~A., {Danforth}, C.~W., {et~al.} 2013, \apj, 763,
  148

\bibitem[{{Sutherland} \& {Dopita}(1993)}]{sd93}
{Sutherland}, R.~S., \& {Dopita}, M.~A. 1993, \apjs, 88, 253

\bibitem[{{Thom} \& {Chen}(2008)}]{ThomChen_2008ApJS..179...37T}
{Thom}, C., \& {Chen}, H.-W. 2008, \apjs, 179, 37

\bibitem[{{Tripp} {et~al.}(2008){Tripp}, {Sembach}, {Bowen}, {Savage},
  {Jenkins}, {Lehner}, \& {Richter}}]{Tripp_2008ApJS..177...39T}
{Tripp}, T.~M., {Sembach}, K.~R., {Bowen}, D.~V., {et~al.} 2008, \apjs, 177, 39

\bibitem[{{Tumlinson} {et~al.}(2011){Tumlinson}, {Thom}, {Werk}, {Prochaska},
  {Tripp}, {Weinberg}, {Peeples}, {O'Meara}, {Oppenheimer}, {Meiring}, {Katz},
  {Dav{\'e}}, {Ford}, \& {Sembach}}]{Tumlinson_2011Sci...334..948T}
{Tumlinson}, J., {Thom}, C., {Werk}, J.~K., {et~al.} 2011, Science, 334, 948

\bibitem[{{Viel} {et~al.}(2017){Viel}, {Haehnelt}, {Bolton}, {Kim}, {Puchwein},
  {Nasir}, \& {Wakker}}]{Viel_2017MNRAS.467L..86V}
{Viel}, M., {Haehnelt}, M.~G., {Bolton}, J.~S., {et~al.} 2017, \mnras, 467, L86

\bibitem[{{Voit}(2018)}]{Voit_2018arXiv180306036V}
{Voit}, G.~M. 2018, ArXiv e-prints, arXiv:1803.06036

\bibitem[{{Voit} {et~al.}(2015{\natexlab{a}}){Voit}, {Bryan}, {O'Shea}, \&
  {Donahue}}]{Voit_PrecipReg_2015ApJ...808L..30V}
{Voit}, G.~M., {Bryan}, G.~L., {O'Shea}, B.~W., \& {Donahue}, M.
  2015{\natexlab{a}}, \apjl, 808, L30

\bibitem[{{Voit} {et~al.}(2015{\natexlab{b}}){Voit}, {Donahue}, {Bryan}, \&
  {McDonald}}]{Voit_2015Natur.519..203V}
{Voit}, G.~M., {Donahue}, M., {Bryan}, G.~L., \& {McDonald}, M.
  2015{\natexlab{b}}, \nat, 519, 203

\bibitem[{{Voit} {et~al.}(2015{\natexlab{c}}){Voit}, {Donahue}, {O'Shea},
  {Bryan}, {Sun}, \& {Werner}}]{Voit+2015ApJ...803L..21V}
{Voit}, G.~M., {Donahue}, M., {O'Shea}, B.~W., {et~al.} 2015{\natexlab{c}},
  \apjl, 803, L21

\bibitem[{{Voit} {et~al.}(2005){Voit}, {Kay}, \& {Bryan}}]{vkb05}
{Voit}, G.~M., {Kay}, S.~T., \& {Bryan}, G.~L. 2005, \mnras, 364, 909

\bibitem[{{Voit} {et~al.}(2018){Voit}, {Ma}, {Greene}, {Goulding}, {Pandya},
  {Donahue}, \& {Sun}}]{Voit2018_LX-T-R}
{Voit}, G.~M., {Ma}, C.~P., {Greene}, J., {et~al.} 2018, \apj, 853, 78

\bibitem[{{Voit} {et~al.}(2017){Voit}, {Meece}, {Li}, {O'Shea}, {Bryan}, \&
  {Donahue}}]{Voit_2017_BigPaper}
{Voit}, G.~M., {Meece}, G., {Li}, Y., {et~al.} 2017, \apj, 845, 80

\bibitem[{{Werk} {et~al.}(2014){Werk}, {Prochaska}, {Tumlinson}, {Peeples},
  {Tripp}, {Fox}, {Lehner}, {Thom}, {O'Meara}, {Ford}, {Bordoloi}, {Katz},
  {Tejos}, {Oppenheimer}, {Dav{\'e}}, \& {Weinberg}}]{Werk_2014ApJ...792....8W}
{Werk}, J.~K., {Prochaska}, J.~X., {Tumlinson}, J., {et~al.} 2014, \apj, 792, 8

\bibitem[{{Werk} {et~al.}(2016){Werk}, {Prochaska}, {Cantalupo}, {Fox},
  {Oppenheimer}, {Tumlinson}, {Tripp}, {Lehner}, \&
  {McQuinn}}]{Werk2016_ApJ...833...54W}
{Werk}, J.~K., {Prochaska}, J.~X., {Cantalupo}, S., {et~al.} 2016, \apj, 833,
  54

\bibitem[{{Yoshino} {et~al.}(2009){Yoshino}, {Mitsuda}, {Yamasaki}, {Takei},
  {Hagihara}, {Masui}, {Bauer}, {McCammon}, {Fujimoto}, {Wang}, \&
  {Yao}}]{Yoshino_2009PASJ...61..805Y}
{Yoshino}, T., {Mitsuda}, K., {Yamasaki}, N.~Y., {et~al.} 2009, \pasj, 61, 805

\bibitem[{{Zahedy} {et~al.}(2018){Zahedy}, {Chen}, {Johnson}, {Pierce},
  {Rauch}, {Huang}, {Weiner}, \& {Gauthier}}]{Zahedy_2018arXiv180905115Z}
{Zahedy}, F.~S., {Chen}, H.-W., {Johnson}, S.~D., {et~al.} 2018, ArXiv
  e-prints, arXiv:1809.05115

\bibitem[{{Zhu} {et~al.}(2014){Zhu}, {M{\'e}nard}, {Bizyaev}, {Brewington},
  {Ebelke}, {Ho}, {Kinemuchi}, {Malanushenko}, {Malanushenko}, {Marchante},
  {More}, {Oravetz}, {Pan}, {Petitjean}, \&
  {Simmons}}]{Zhu_2014MNRAS.439.3139Z}
{Zhu}, G., {M{\'e}nard}, B., {Bizyaev}, D., {et~al.} 2014, \mnras, 439, 3139

\end{thebibliography}

\end{document}